\documentclass[aps,pra,nofootinbib,twocolumn,floatfix,showpacs,superscriptaddress,longbibliography]{revtex4-1}
\usepackage[utf8]{inputenc}
\usepackage{amsmath,graphicx,epsfig,amsthm,color}
\usepackage{pifont,bbding,amssymb,soul}
\usepackage{indentfirst}
\usepackage{times}
\usepackage{appendix, comment}
\usepackage{multirow}
\usepackage{makecell}

\usepackage{tikz,pgfplots}

\newcommand{\proj}[1]{\ket{#1}\!\bra{#1}}
\newcommand{\tr}{{\rm tr}}

\newcommand{\Iff}{\emph{iff} }

\newcommand{\vcr}{v_{\mathrm{cr}}}
\newcommand{\od}{\frac{1}{d}}

\newcommand{\F}{\mathcal{F}}
\newcommand{\Q}{\mathcal{Q}}
\newcommand{\bra}[1]{\langle #1|}
\newcommand{\ket}[1]{|#1\rangle}
\newcommand{\bracket}[2]{\langle #1|#2\rangle}

\newcommand{\mes}{\ket{\Phi^+_d}}
\newcommand{\pmes}{\proj{\Phi^+_d}}

\newcommand{\rhoa}{\rho_{\mathrm{ave}}}
\newcommand{\rhof}{\rho_f}
\newcommand{\rhos}{\rho_{\mathrm{sep}}}
\newcommand{\rhoq}{\rho(q)}
\newcommand{\rhofk}{\rho_{f,\kappa}(q)}
\newcommand{\rhofkp}{\rho_{f,\kappa'}(q)}
\newcommand{\rhoW}{W(v)}
\newcommand{\rhoWf}{W_f(v)}
\newcommand{\rin}{\varrho}

\newcommand{\Ak}{A_{K}}
\newcommand{\pW}{p_\text{\tiny W}(v)}

\newcommand{\pk}{p_{\kappa}(q)}
\newcommand{\pkp}{p_{\kappa'}(q)}

\newcommand{\fc}{f_{\mathrm{c}}}
\newcommand{\Fc}{F_{\mathrm{c}}}

\newcommand{\id}{\mathbb{I}}

\usepackage[hyperindex,breaklinks,colorlinks=true,citecolor=blue,urlcolor=blue]{hyperref}

\usepackage[capitalize]{cleveref}

\begin{document}

\title{Activating Hidden Teleportation Power: Theory and Experiment}

\author{Jyun-Yi Li}
\email{These authors contributed equally to this work}
\affiliation{Department of Physics and Center for Quantum Frontiers of Research \& Technology (QFort), National Cheng Kung University, Tainan 701, Taiwan}

\author{Xiao-Xu Fang}
\email{These authors contributed equally to this work}
\affiliation{School of Physics, Shandong University, Jinan 250100, China}

\author{Ting Zhang}
\affiliation{School of Physics, Shandong University, Jinan 250100, China}

\author{Gelo Noel M. Tabia}
\email{tgnm@mx.nthu.edu.tw}
\affiliation{Department of Physics and Center for Quantum Frontiers of Research \& Technology (QFort), National Cheng Kung University, Tainan 701, Taiwan}
\affiliation{Center for Quantum Technology, National Tsing Hua University, Hsinchu 300, Taiwan}

\author{He Lu}
\email{luhe@sdu.edu.cn}
\affiliation{School of Physics, Shandong University, Jinan 250100, China}

\author{Yeong-Cherng Liang}
\email{ycliang@mail.ncku.edu.tw}
\affiliation{Department of Physics and Center for Quantum Frontiers of Research \& Technology (QFort), National Cheng Kung University, Tainan 701, Taiwan}
\affiliation{Physics Division, National Center for Theoretical Sciences, Taipei 10617, Taiwan}

\date{\today}

\begin{abstract}
Ideal quantum teleportation transfers an unknown quantum state intact from one party Alice to the other Bob via the use of a maximally entangled state and the communication of classical information. If Alice and Bob do not share entanglement, the maximal average fidelity between the state to be teleported and the state received, according to a classical measure-and-prepare scheme, is upper bounded by a function $\fc$ that is inversely proportional to the Hilbert space dimension. In fact, even if they share entanglement, the so-called teleportation fidelity may still be less than the classical threshold $\fc$. For two-qubit entangled states, conditioned on a successful local filtering, the teleportation fidelity can always be activated, i.e., boosted beyond $\fc$. Here, for {\em all} dimensions larger than two, we show that the teleportation power {\em hidden} in a subset of entangled two-qudit Werner states can also be activated. In addition, we show that an entire family of two-qudit rank-deficient states violates the reduction criterion of separability, and thus their teleportation power is either above the classical threshold or can be activated. Using hybrid entanglement prepared in photon pairs, we also provide the first proof-of-principle experimental demonstration of the activation of teleportation power hidden in this latter family of qubit states. The connection between the possibility of activating hidden teleportation power with the closely-related problem of entanglement distillation is discussed.
\end{abstract}

\maketitle

\section{Introduction}

In quantum information science, entanglement~\cite{Horodecki:RMP:2009} serves as a resource within the paradigm of local operations assisted by classical communications (LOCC). In fact, sharing entanglement is essential for exhibiting a quantum advantage over classical resources in computation~\cite{rspa.2002.1997,Vidal:PRL:2003}, secret key distribution~\cite{Ekert91}, superdense coding~\cite{PhysRevLett.69.2881}, and metrology~\cite{Toth:JPA:2014}, etc. Among the  possibilities that entanglement empowers, quantum teleportation~\cite{teleportation}, i.e., the transfer of quantum states using shared entanglement and classical communication, is especially worth noting (see, e.g.,~\cite{Pirandola2015, Qfundation} for some recent advances).

Indeed, teleportation serves as a primitive in various quantum protocols such as remote state preparation~\cite{StatePreparation,Devetak:PRL:2001},  entanglement swapping~\cite{Zukowski:PRL:1993}, and quantum repeaters~\cite{QuantumRepeaters}. In universal quantum computing with linear optics, it enables near-deterministic two-qubit gates ~\cite{Gottesman1999} and makes assembling cluster states more efficient~\cite{PhysRevLett.93.040503, PhysRevLett.95.010501}. Theoretically, it has been used as a tool for exploring closed timelike curves~\cite{PhysRevLett.106.040403} and black hole evaporation~\cite{Lloyd2014}. Recently, it was used to experimentally demonstrate the scrambling of quantum information~\cite{Landsman2019}.  In this work, we compare entangled states to {\em classical} resources for the task of teleportation.

In the original protocol~\cite{teleportation}, two remote parties (called Alice and Bob) share an entangled pair of qubits. By performing a joint measurement on her half of the entangled qubit and an unknown qubit $\ket{\psi}$ given to her, Alice teleports $\ket{\psi}$ to Bob by transmitting only the classical measurement outcome to Bob. The quality of this state transfer is quantified~\cite{PopescuTB} by the \emph{teleportation fidelity}~\cite{Jozsa,Liang_2019}, which measures the {\em average} overlap between $\ket{\psi}$ and the state Bob receives.

To teleport a quantum state perfectly, sharing a maximally entangled state is imperative. However, due to decoherence, this ideal resource is often not readily shared between remote parties, thus resulting in a non-ideal teleportation fidelity. When the entanglement is too weak, the  teleportation fidelity can even be simulated by adopting a measure-and-prepare scheme~\cite{PopescuTB}, without sharing any entanglement. Thus, whenever an entangled state yields a teleportation fidelity larger than the classical threshold of  $\fc=\frac{2}{d+1}$~\cite{Horodecki}, it is conventionally said to be useful for teleportation, but otherwise useless  (see~\cite{Cavalcanti:PRL:2017,arXiv:2007.04658} for some other notions of non-classicality). Here, $d$ is  the local state space dimension.

Importantly, teleportation power, as with some other desirable features of an entangled state, may  be activated by utilizing experimentally-feasible~\cite{ExpHiddenNonlocality,Pramanik:PRA:2019,Nery:PRL:2020} {\em local filtering}~\cite{Gisin} operations (see also~\cite{PopescuHN,Peres:PRA:1996,Masanes,Masanes:PRL:2008,Liang:PRA:2012}). Accordingly, we say that $\rho$ has {\em hidden teleportation power} (HTP) if it is {\em useless} for teleportation but becomes {\em useful}, i.e., {\em activated} after a successful local filtering.  Two-qubit entangled states are either useful or can be activated~\cite{Horodecki97,Verstraete:PRL:2003,Horodecki00}.  Bound entangled~\cite{Horodeck:PRL:1998} states are useless for teleportation and cannot be activated~\cite{Horodecki} while {\em all} entangled isotropic states~\cite{ReductionCriterion} are useful. Are there higher-dimensional entangled states whose teleportation power can be activated? Here, we show  that for {\em all} dimensions $d\ge 3$, entangled Werner states~\cite{Werner:PRA:1989} exhibiting HTP can be found. Moreover, a family of rank-deficient states is provably useful or can have its teleportation power activated. We further provide the {\em first} proof-of-principle experimental demonstration of this activation process using entangled photon pairs, pushing the frontier of photonics teleportation experiments (see, e.g.,~\cite{Bouwmeester:1997aa,Boschi1998,Pan:PRL:1998,Furusawa706,Marcikic:2003aa,Ma:2012aa,Jin2010,Jiang2019,Luo:PRL:2019,Hu:PRL:2020}) in another direction.

For {\em any} two-qudit state $\rho$, determining its teleportation fidelity $f_d(\rho)$ and hence its usefulness is {\em a priori} not trivial as this requires an integration over all pure states $\ket{\psi}$ chosen uniformly from $\mathbb{C}^d$. However, $f_d(\rho)$ is known~\cite{Horodecki} to relate monotonically to the fully entangled fraction (FEF) of $\rho$, denoted by $F_d(\rho)$ as
\begin{equation}
	f_d(\rho)=\frac{F_{d}(\rho)d+1}{d+1},\quad
	F_{d}(\rho)=\max_{\ket{\Psi_d}} \bra{\Psi_d}\rho\ket{\Psi_d}
\label{eqn:FEF}
\end{equation}
where
\begin{equation}\label{Eq:Psi_d}
	\ket{\Psi_d} = (\id_d \otimes U_d) \mes
\end{equation}
is an arbitrary $d$-dimensional maximally entangled state, $\id_d$ is the $d\times d$ identity matrix, $U_d$ is a $d\times d$ unitary matrix and $\mes =\frac{1}{\sqrt{d}}\sum_{i=0}^{d-1}\ket{i}\ket{i}$. The classical measure-and-prepare  threshold $\fc=\frac{2}{d+1}$ corresponds to an FEF of $\Fc=\od$.  Hence, a quantum state $\rho$ is {\em useful} for teleportation if and only if ({$\Iff$})  $F_{d}(\rho)>\Fc$.
\newline

\section{Boosting teleportation power}
We are interested in activating the usefulness for teleportation by local filtering (i.e., stochastic LOCC~\cite{Vidal:JMO:2000}). Formally, local filtering on a bipartite system $\rho$ gives $\tau=(A\otimes B)\rho(A\otimes B)^{\dagger}$, where the filters $A$ and $B$ are $d \times d$ matrices having bounded singular values. Through renormalization, we may set $||A||_{\infty}=||B||_{\infty}= 1$, i.e., their largest singular value being unity. Conditioned on successful filtering, which happens with probability $p=\mathrm{tr}(\tau)$, the resulting filtered state is $\rhof=\frac{\tau}{p}$.  Generally, a trade-off between the maximization of $F_d[\rhof]$ and the corresponding success probability is expected.

Physically relevant filtering should give $p\neq 0$. Then, the process of boosting teleportation power can be made deterministic~\cite{Verstraete:PRL:2003} by preparing a separable state, say, $\rhos=\ket{\phi}\ket{\varphi}\bra{\phi}\bra{\varphi}$ whenever the filtering operation fails. Explicitly, this {\em average} state
$\rhoa=p\rhof+(1-p)\rhos$
can be obtained as the output of the completely-positive trace-preserving map
\begin{equation}
\rhoa=M_{1}\rho M_{1}^{\dagger}+\sum_{i,j,k}M_{ijk}\rho M_{ijk}^{\dagger}
\end{equation}
where the Kraus operator  $M_{ijk}=\ket{\phi}\ket{\varphi}\bra{i}\bra{j}G_{k}\ (i, j=0, 1,..., d-1)$, $M_1=A\otimes B$,  with  $G_1=\sqrt{\id_d-A^{\dagger}A}\otimes \sqrt{\id_d-B^{\dagger}B}$,
$G_2=A\otimes \sqrt{\id_d-B^{\dagger}B}$, and $G_3=\sqrt{\id_d-A^{\dagger}A}\otimes B$.

\subsection{Deterministic teleportation protocol with filtering}
\label{Sec:DeterministicProtocol}

Consequently, the teleportation protocol can also be made deterministic by incorporating the various outcomes of the local filtering process.  For simplicity, the following discussion assumes that Alice (the sender) and Bob (the receiver) share a two-qubit entangled state $\rho_{AB}$ and where the unknown state to be teleported $\ket{\psi}_{T}$ is also a qubit. The protocol can be straightforwardly generalized to the case involving higher-dimensional quantum states.

\begin{enumerate}
    \item
    First,  Bob applies his local filter on qubit $B$.
    He then sends a bit $b$ to Alice to inform her whether the filtering succeeded ($b=1$) or failed ($b=0$).

    \item
    \begin{enumerate}
    	\item If $b=1$, Alice performs a local filtering operation on qubit $A$.
	And If her filtering succeeds, Alice performs a Bell-state measurement on the qubit pair $(T,A)$ and sends the two-bit measurement outcome $(ij) \in \{ 00,01,10,11 \}$ to Bob.
    \item If $b=0$ or if her filtering operation fails, Alice measures qubit $T$ in the computational basis and sends her measurement outcome $a = 0,1$ (corresponding to $\ket{0}$ and $\ket{1}$) to Bob.
    \end{enumerate}

    \item Depending on the number of bits he receives, Bob knows if Alice's local filtering succeeded. He then acts accordingly to complete the teleportation protocol.
    \begin{enumerate}
    	\item If Bob receives one bit $a$, he locally prepares the computational basis state $\ket{a}$.
	\item If Bob receives two bits $(ij)$, he applies the unitary (Pauli) correction $Z^{i}X^{j}$ on his qubit $B$.
    \end{enumerate}

\end{enumerate}

The output of the protocol is Bob's final qubit. If any local filtering fails, it would be a qubit prepared in some computational basis state $\ket{a}$, which always contributes $\frac{1}{d}$ to the fully entangled fraction. Otherwise, it will be the unitarily-corrected qubit from Bob's share of $\rho_{AB}$.

\subsection{Figures of merit}

There are thus two natural figures of merit relevant to boosting the teleportation power of $\rho$. The first of these concerns
\begin{equation}\label{Eq:MaxFEFFiltered}	
\begin{split}
	\qquad \max_{A,B}\quad &F_{d}[\rhof(A,B)] = \max_{A,B}\quad \bra{\Phi_d^+}\rhof(A,B)\ket{\Phi_d^+},\\
	\text{such that \ } &\qquad||A||_{\infty}= 1,\quad ||B||_{\infty}= 1
\end{split}
\end{equation}
where the equality in the objective function follows by absorbing the $U_d$ defining $F_d$ [\cref{Eq:Psi_d}] into the definition of Bob's filter $B$. Consequently, in maximizing  the {\em alternative} figure of merit $K(\rho)\equiv F_d(\rhoa)$, called the cost function in~\cite{Verstraete:PRL:2003}, one may set $\rhos=\ket{0}\ket{0}\bra{0}\bra{0}$, thus giving
\begin{equation}\label{Eq:CostFunction-Simplified}
	K(\rho) = p F_{d}(\rhof)+\frac{1-p}{d},
\end{equation}
which exceeds $\frac{1}{d}$ \Iff $F_d(\rhof)>\frac{1}{d}$.  Note that the deterministic teleportation protocol described in~\cref{Sec:DeterministicProtocol} ensures that the cost function of~\cref{Eq:CostFunction-Simplified} is attained.

Hence, although the optimal filter(s) and the final FEF may depend on the choice among these figures of merit,  the possibility of activating $\rho$ does not. That is, $\rho$ displays HTP \Iff it satisfies two {\bf conditions}:\begin{align*}
&\text{\bf condition } (\mathsf{a}):\quad F_d(\rho)\le \frac{1}{d}, \text{ and}\\
&\text{\bf condition } (\mathsf{b}):\quad F_d(\rhof) \text{ or } F_d(\rhoa)>\frac{1}{d}.
\end{align*}
Condition ($\mathsf{a}$) induces~\cite{Ganguly:PRL:2011,Zhao:PRA:2012} a convex set and qualifies the {\em uselessness}~\cite{PopescuHN,Horodecki} of $\rho$ for teleportation but the set of $\rho$ complying with ($\mathsf{b}$) is concave.

Two {\bf facts} about the reduction criterion of separability (RC)~\cite{ReductionCriterion} should now be noted:
\begin{enumerate}
\item[(I)]\label{RC:NV} the non-violation of RC by $\rho$ guarantees condition ($\mathsf{a}$)
\item[(II)]\label{RC:V} the violation of RC by $\rho$ implies ($\mathsf{b}$) even with single-side filtering.
\end{enumerate}
However, there seems to be no single figure of merit fully characterizing both conditions simultaneously.

\subsection{Werner states}

Consider the Werner state~\cite{Werner:PRA:1989}:
\begin{equation}
\rhoW=\frac{2v}{d(d+1)}P_{+}+\frac{2(1-v)}{d(d-1)}P_{-}, \quad v\in[0,1]
\end{equation}
where $P_{\mp}=(\id_{d^2}\mp V)/2$ is the projector onto the (anti)symmetric subspace of $\mathbb{C}^d\otimes\mathbb{C}^d$ and $V=\sum^{d-1}_{i,j=0}\ket{i}\ket{j}\bra{j}\bra{i}$ is the swap operator. $\rhoW$ is entangled \Iff $0\le v<\frac{1}{2}$. For $d>2$,  {\em all}  $\rhoW$ satisfy~\cite{ReductionCriterion} RC and thus, by {\bf fact} (I), are useless for teleportation. However,  as shown below, the teleportation power of some entangled $\rhoW$ can be activated.

Specifically, we perform optimizations of~\cref{Eq:MaxFEFFiltered} using the MATLAB function $\mathsf{fminunc}$ with more than $3\times10^4$ random initial parameters for both $d=3,4$, in addition to several optimizations for larger values of $d$. Let $\rhoWf$ be the state filtered from $\rhoW$. The largest value of $F_d[\rhoWf]$ we found happens to be attainable with the {\em qubit} filters
\begin{equation}
A_W = \sigma_z\oplus \mathbf{0}_{d-2},\quad B_W=\sigma_x\oplus \mathbf{0}_{d-2},
\end{equation}
where $\sigma_{x},\sigma_{z}$ are Pauli matrices, and $\oplus\, \mathbf{0}_{D}$ means a direct sum with a zero matrix of size $D$.  The filtering succeeds with probability $\pW=\frac{2N}{d(d^2-1)}$ where $N=(d+1)(1-v)+3v(d-1)$. Interestingly, even if we take into account $\pW$ and maximize the cost-function $K[\rhoW]$, the best filters found remain unchanged.

\begin{figure}
\includegraphics[]{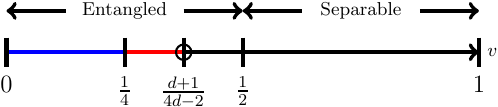}
\caption{The Werner state $\rhoW$, $v\in[0,1]$ is entangled \Iff   $0\le v<\frac{1}{2}$. We show in this work  that for $d>2$, $\rhoW$ has HTP in the region $0\le v< \frac{d+1}{4d-2}$ (blue and red segment). As $d$ increases from 2 towards $\infty$, the threshold $\vcr=\frac{d+1}{4d-2}$ moves from $\frac{1}{2}$ towards $\frac{1}{4}$, as symbolized by the (shrinking) red segment. The blue segment indicates the region where $\rhoW$ always has HTP whenever $d>2$. For $d=2$, all entangled $\rhoW$ are useful for teleportation.}
\label{Fig:NumberLine}
\end{figure}

For activation, locally-filtering $\rhoW$ onto the {\em same} qubit subspace suffices.  However, with the Pauli rotations, $\rhoWf$ takes the simple  form
\begin{align}
\nonumber
\rhoWf&=\frac{1}{N}[(d+1)(1-v)\proj{\Phi^+_2} \\
 &\quad +v(d-1)(\id-\proj{\Phi^+_2}]\oplus \mathbf{0}_{d^2-4}.
\end{align}
Its FEF $F_{d}[\rhoWf]=\frac{2(d+1)(1-v)}{dN}$ beats the classical limit  $F_c=\frac{1}{d}$ whenever $v<\vcr=\frac{d+1}{4d-2}$. Therefore, for $d>2$, $\rhoW$ exhibits HTP for $0\le v<\vcr$, as shown in Fig.~\ref{Fig:NumberLine}. For completeness, we illustrate in~\cref{App:Werner} how the increase in FEF, i.e., $F_{d}[\rhoWf]-F_{d}[\rhoW]$  varies with the success probability $\pW$ of filtering.

Evidently, qubit filters  introduce asymmetries by favoring a 2-dimensional subspace of $\mathbb{C}^d$ while giving a poor fidelity when teleporting any $\ket{\psi}$ lying in the complementary subspace. However, since the set of $\ket{\psi}\in\mathbb{C}^{d'}$ with $d'<d$ constitute a set of measure zero in $\mathbb{C}^d$, these asymmetric fidelities do not contribute to the computation of the teleportation fidelity $f_d(\rho)$, which averages over all $\ket{\psi}\in\mathbb{C}^d$. Still, it seems intriguing that such filters optimize~\cref{Eq:MaxFEFFiltered}, as our numerical results suggest.

\subsection{Rank-deficient states}

Next, consider a family of two-qudit, rank-two entangled states~\cite{Horodecki,Verstraete:PRL:2003}
\begin{equation}\label{Eq:rhoq}
	\rho(q) = q \pmes + (1-q)\proj{0}\otimes\proj{1}, \quad q\in(0,1]
\end{equation}
where $q\in(0,1]$. Throughout, we shall only state our findings while leaving all technical details to the Appendices. Firstly, $\rhoq$ is provably (see~\cref{App:FEF:Rhoq}) useful for teleportation for {\em all} $d\ge 4$, but for $d\le3$, only when  $q>\frac{1}{d}$.

By identifying an eigenvector with negative eigenvalue, we further show in \cref{App:VReduction} that $\rhoq$ {\em violates} the RC:
\begin{equation}\label{Eq:RCA}
    \tr_B[\rhoq]\otimes \id_d - \rhoq \succeq 0
\end{equation}
where $\tr_B(\cdot)$ denotes the partial trace over $B$ and $\succeq 0$ means matrix positivity.

Thus, by {\bf fact} (II), filtering on one side (Alice) guarantees that the FEF of $\rhoq$ can be boosted beyond $F_c$.  In this case, the filter maximizing $K[\rho(q)]$, as we show in~\cref{App:OptimalOneSide}, is
\begin{equation}\label{Eq:OptimalFilterRhoq}
	\Ak = \kappa\proj{0} + \sum_{j=1}^{d-1} \proj{j},
\end{equation}
where $\kappa=\frac{(d-1)q}{d(1-q)}$.  This reduces to the optimal filter found~\cite{Verstraete:PRL:2003} in the $d=2$ case. Our numerical results obtained by maximizing $K[\rho(q)]$ suggest that $\Ak$  may even be optimal when two-side filtering is allowed.\footnote{Note that filters giving $F_d[\rhof(q)]\to 1$ but with vanishing success probability are known~\cite{Horodecki}. See also~\cref{App:TwoSide}.}

\begin{figure}[t!bp]
  \includegraphics[scale=1]{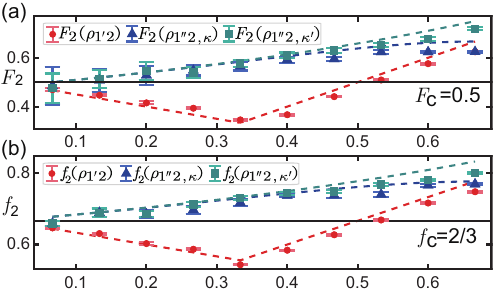}
 \caption{
 Theoretical (dashed lines) and experimental (markers) results illustrating the teleportation power before and after filtering for qubit $\rhoq$.  {\bf (a).} FEF $F_2$ [evaluated using~\cref{eqn:FEF}] {\bf (b).} teleportation fidelity $f$. In each plot, the bottom (red) results correspond to the unfiltered states, i.e., $\rhoq$ (theory) and  $\rho_{1'2}$ (experiment) in~\cref{Fig:exp_setup}(a). The middle (blue) results are for the filtered state $\rhofk$ (theory) and $\rho_{1''2,\kappa}$ (experiment) in~\cref{Fig:exp_setup}(a) while the top (turquoise) results are for the filtered state $\rhofkp$ (theory) and $\rho_{1''2,\kappa'}$ (experiment) in~\cref{Fig:exp_setup}(a). HTP is shown if a red marker is below the solid line but the corresponding blue or turquoise marker is above the same line, which happens only when $q\in(0, \frac{1}{2}]$.
}
\label{Fig:exp_data}
 \end{figure}

Let us define the subnormalized state $\ket{\chi}:= \frac{1}{\sqrt{d}}(\kappa\ket{0}\ket{0} + \sum_{i=1}^{d-1} \ket{i}\ket{i})$.
Then, conditional upon a successful filtering, which occurs with probability
\begin{equation}\label{Eq:pk}
	\pk=\kappa^2 \left(\frac{q}{d}+1-q \right)+\frac{q}{d}(d-1),
\end{equation}
one obtains the filtered state
\begin{equation}\label{eqn:rhof}
	\rhofk = \frac{1}{\pk}(q \proj{\chi} + (1-q)\kappa^2\proj{0}\otimes\proj{1}),
\end{equation}
which has an FEF of:
\begin{equation}\label{eqn:FK}
	F_{d}[\rhofk]=\tfrac{q}{d^{2}\pk}\left(\kappa+d-1\right)^2,\,\, q\in\left(0, \tfrac{d}{2d-1}\right).
\end{equation}
That this shows the HTP of certain $\rho(q)$ is illustrated for the qubit case in \cref{Fig:exp_data} (and in~\cref{App:TwoSide} for the qutrit case). More generally, for all $d\ge 2$, one finds an increase in FEF, i.e., $F_d[\rhofk]>F_d[\rho(q)]$ for $q\in\left(0, \tfrac{d}{2d-1}\right)$.

For comparison, we also compute~\cref{Eq:MaxFEFFiltered} by filtering only on Alice's side. In this case, our numerical results suggest that the best local filter  takes the same form as $\Ak$ but with $\kappa$ replaced by $\kappa'=\frac{q}{q+d(1-q)}$. The success probability $\pkp$ and the FEF of the filtered state $F_{d}[\rhofkp]$ are analogously obtained by replacing $\kappa$ with $\kappa'$ in \cref{eqn:FK}.
\newline

\section{Experimental demonstration}

\begin{figure*}[t!bp]
 \includegraphics[]{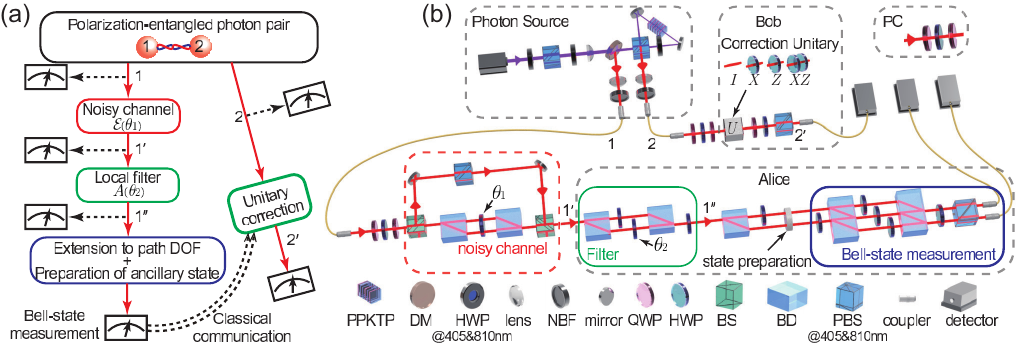}
 \caption{
 {\bf (a)}, Experimental scheme used in demonstrating the HTP of $\rhoq$. QST may be performed at stage $1$, $1^{\prime}$, and $1^{\prime\prime}$ to estimate the density matrix corresponding, respectively, to the initial entangled state $\rho_{12}$, the experimentally prepared state $\rho_{1^\prime2}$ [for $\rhoq$], and the locally filtered state $\rho_{1^{\prime\prime}2,\kappa}$ or $\rho_{1^{\prime\prime}2,\kappa'}$ [for $\rhofk$ and $\rhofkp$]. {\bf (b)},  Experimental setup. The generated entangled photons are each coupled into a single-mode fiber and sent to Alice and Bob via optical fibers. The fiber-induced polarization drift is corrected by a polarization controller (PC), which is a half-wave plate (HWP) sandwiched by two quarter-wave plates (QWPs). The noisy channel generates $\rhoq$ according to $\theta_1$, which is then filtered to boost its teleportation power. In our setup, the classical communication was carried out after the experiment, i.e., Bob applies the unitary correction on photon 2$^{\prime}$ in a post-selected manner to recover the teleported state. See text and~\cref{App:Exp} for details. DM: dichroic mirror. NBF: narrow-band filter. BS: beam splitter.
}
 \label{Fig:exp_setup}
 \end{figure*}

Experimentally, we prepare two-qubit $\rho(q)$ for $q=\frac{1}{15}, \frac{2}{15},\ldots, \frac{10}{15}$ and demonstrate how one-side local filtering can be applied to boost its teleportation power.  \cref{Fig:exp_setup}(a) summarizes our  protocol and \cref{Fig:exp_setup}(b) shows the  experimental setup (with no measurement  at stage 1,1$^\prime$, nor $1^{\prime\prime}$). Polarization-entangled photon pairs are first generated via a periodically-poled potassium titanyl phosphate (PPKTP) crystal in a Sagnac interferometer~\cite{Kim06}, which is bidirectionally pumped by a 405nm ultraviolet diode laser. From quantum state tomography (QST), we estimate that the generated entangled state $\rho_{12}$ has a fidelity of $0.954\pm0.003$ with  $\ket{\Psi_2^{+}}_{12}=\frac{1}{\sqrt{2}}(\ket{H}\ket{V}+\ket{V}\ket{H})$, where the $H$ (horizontal) and $V$ (vertical) polarization encode, respectively, $\ket{0}$ and $\ket{1}$.
QST requires both photons to be measured in different bases, which we achieve by passing them through wave plates with the appropriate setting and a polarizing beam splitter (PBS) before detection.

To generate $\rhoq$,  we let the photon pass through a noisy channel $\mathcal{E}(\theta_1)$, see~\cref{Fig:exp_setup}(a), such that $\rho_{1^\prime2}(\theta_1)=\mathcal{E}(\theta_1)\otimes
\id_2(\ket{\Psi_2^{+}}\bra{\Psi_2^{+}})=q(\theta_1)\ket{\Phi_2^{+}}\bra{\Phi_2^{+}}+[1-q(\theta_1)]\ket{H}\ket{V}\bra{H}\bra{V}$.
The parameter $q(\theta_1)=\frac{2\sin^2(2\theta_1)}{1+2\sin^2(2\theta_1)}$ is varied by rotating the angle $\theta_1$ of the  HWP between the two beam displacers (BDs). To determine the FEF before filtering, we estimate $\rho_{1^\prime2}$ from QST and compute \cref{eqn:FEF}. The results are shown as red dots in ~\cref{Fig:exp_data}(a). For $q(\theta)\le \frac{7}{15}$, we observe that $F_2(\rho_{1^\prime2})<\Fc=\frac{1}{2}$, thus certifying their uselessness for teleportation.

To boost their teleportation power, we apply filter $\Ak=\mathrm{diag}[\kappa,1]$ on photon $1^{\prime}$ by implementing an amplitude damping channel~\cite{Nielsen10} and keeping only the photons exiting from one specific output port~\cite{Fisher2012}. The parameter $\kappa$ is related to the angle $\theta_2$ of the HWP at the lower arm by $\kappa=\sin2\theta_2$. By setting  $\sin2\theta_2=\frac{q(\theta_1)}{2[1-q(\theta_1)]}$, we realize the filters $\Ak$ with parameter $\kappa$ and obtain $\rho_{1^{\prime\prime}2,\kappa}$ with a success probability of $\pk$. Similarly, by tuning $\theta_2$, we can implement the filter $A_{\kappa'}$ and obtain $\rho_{1^{\prime\prime}2,\kappa^{\prime}}$. $F_2(\rho_{1^{\prime\prime}2,\kappa^{\prime}})$ is then similarly estimated.

From~\cref{Fig:exp_data}(a),  we see that except for $q=\frac{1}{15}$, both $F_2(\rho_{1^{\prime\prime}2,\kappa})$ and $F_2(\rho_{1^{\prime\prime}2,\kappa'})$ exceed $\Fc=\frac{1}{2}$ after local filtering, confirming that the filtered states $\rho_{1^{\prime\prime}2,\kappa}$ and $\rho_{1^{\prime\prime}2,\kappa'}$ possess  teleportation power that outperforms the classical measure-and-prepare strategy. To better understand how the filtered states fare in an actual teleportation experiment, we skip the QST for photon 1 (see~\cref{Fig:exp_setup}(b)) for some of the runs and follow the two-photon teleportation scheme of~\cite{Boschi1998} (see also~\cite{Jin2010,Jiang2019}) to provide a proof-of-principle demonstration of activation. In particular, we  introduce a third qubit by involving also the path degree of freedom of photon $1^{\prime\prime}$ at the state-preparation stage in~\cref{Fig:exp_setup}(b).

To verify the teleportation power of the filtered states, we choose for our teleportation experiments the {\em known} input states: $\ket{\psi}\in\{\ket{0},\ket{1},\ket{+}=\frac{1}{\sqrt{2}}(\ket{0}+\ket{1}, \ket{R}=\frac{1}{\sqrt{2}}(\ket{0}+{\rm i}\ket{1})\}$. Using quantum process tomography (QPT)~\cite{Nielsen10}, we then reconstruct the process matrix $\chi_{\text{exp}}$ of our teleportation channel (see~\cref{App:Process} for details). By definition, the resulting teleportation fidelity $f_2(\rho)$ equals the average identity-gate fidelity $\bar{F}(\rho)$, which relates~\cite{Nielsen2002,OBrien2004} (see also~\cite{Horodecki}) to the process fidelity $\F_p=\tr(\chi_{\text{id}}\chi_{\text{exp}})$
by $\bar{F}(\rho)=[2\F_p(\rho)+1]/3$. Here,
$\chi_{\text{id}}$ is the process matrix of the ideal teleportation channel. Our results plotted at \cref{Fig:exp_data} show that $f_2(\rho)$ shares the same trend as $F_2(\rho)$ when we vary $q(\theta_1)$, thus confirming the linear dependence of $f_2(\rho)$ on $F_2(\rho)$ as required by ~\cref{eqn:FEF}. Deviations from the theoretical predictions are mainly due to higher-order photon-pair production events and misalignment in optimal elements. Further experimental details and  theoretical predictions that fit better with the experimental data can be found in~\cref{App:Exp}.

\section{Discussion}

Incidentally,  the interval of $v$ at which $\rhoW$ exhibits HTP coincides with that where $\rhoW$ is  known to be 1-distillable~\cite{Horodecki:2001:EM:2011326.2011328, PhysRevA.61.062312, PhysRevA.61.062313}. The $n$-distillability problem concerns the conversion of $n\ge 1$ copies of a given state  $\rho$ to a finite number of Bell pairs using LOCC. Since all two-qubit entangled states are distillable~\cite{Horodecki97},  $\rho$ is distillable if there exist qubit projections mapping it to a two-qubit entangled state. With the qubit projection first considered by Popescu~\cite{PopescuHN}, it is known~\cite{Horodecki:2001:EM:2011326.2011328, PhysRevA.61.062312, PhysRevA.61.062313} that $\rhoW$ can be locally filtered to a two-qubit entangled state for  $v\in[0,\vcr)$.

The aforementioned coincidence can thus be appreciated by noting the following {\bf observations}:
\begin{itemize}
\item[(i)] the filtered two-qubit entangled state is locally-equivalent to $\rhoWf$, i.e., an isotropic state~\cite{ReductionCriterion} and hence satisfies $F_2[\rhoWf]>\frac{1}{2}$,
\item[(ii)] {\em any}  two-qubit state $\rhof'$ is easily seen to satisfy $F_2[\rhof']>\frac{1}{2}$ \Iff $F_d[\rhof']>\frac{1}{d}$.
\end{itemize}
Nonetheless, let us remind the reader that the problem of distillation and teleportation-power activation are defined differently. For the 1-distillability of $\rho$ by qubit projection\footnote{General distillation protocols may also involve twirling and other LOCC that cannot be described by local filtering alone.}, one seeks for qubit filters $A$ and $B$ such that $\rhof = \frac{A\otimes B\,\rho\,(A\otimes B)^\dag}{\tr[A\otimes B\,\rho\,(A\otimes B)^\dag]}$ is entangled. However, for the problem of activation, one aims to find filters such that $F_d(\rhof)>\frac{1}{d}$. In particular, optimal filters for the latter problem are generally {\em not} a qubit projection [cf. our example for $\rho(q)$].

Despite this  difference, if $\rho$ is 1-distillable by qubit projection, concatenating this projection with the filters provided in~\cite{Verstraete:PRL:2003} does give a filtered state $\rhof'$ satisfying $F_2[\rhof']>\frac{1}{2}$, and hence $F_d[\rhof']>\frac{1}{d}$ by {\bf observation} (ii) above. Conversely, whenever $F_d(\rhof)>\frac{1}{d}$, we have consistently found (numerically) qubit filter(s) giving a (different)  two-qubit filtered state $\tilde{\rho}_f$ satisfying $F_2(\tilde{\rho}_f)>\frac{1}{2}$. A proof of the implication $F_d(\rhof)>\frac{1}{d}\implies F_2(\tilde{\rho}_f)>\frac{1}{2}$ is, to our knowledge, lacking. If true, then the problem of boosting FEF beyond $F_c$ becomes {\em equivalent} to the problem of 1-distillability by qubit projection, potentially simplifying the analysis of entanglement distillability. Intriguingly, while  qubit filters appear restrictive and introduce asymmetries in teleportation fidelities, they may still guarantee, by {\bf observation} (ii) above, the general usefulness of the filtered state for teleporting a {\em qudit} state. A better understanding of when and why a qubit filter optimizes~\cref{Eq:MaxFEFFiltered} is thus clearly desirable.

On the experimental side, note that a third party Charlie may carry out the state preparation by inserting any combination of wave plates and having them shielded from Alice. This slight modification from our setup allows Alice to teleport~\cite{Boschi1998} {\em any pure state} $\ket{\psi}$ {\em unknown} to her. This and the need to perform a Bell-state measurement distinguish our experiment from that for remote state preparation~\cite{StatePreparation,Peters:PRL:2005}, which only prepares certain {\em known}  states remotely. Nonetheless, if we want to use the filtered state to teleport, e.g., a part of an entangled state (cf. entanglement swapping~\cite{Zukowski:PRL:1993,Pan:PRL:1998}) then we would have to swap an external qubit state with our photon polarization state at stage $1^\prime$. Albeit interesting and relevant, solving this problem is outside the scope of the present proof-of-principle demonstration. An analogous demonstration for higher-dimensional quantum states, given recent progress~\cite{Luo:PRL:2019,Hu:PRL:2020}, would also be timely.

Meanwhile, although~\cite{ExpHiddenNonlocality} experimentally demonstrated {\em hidden nonlocality}~\cite{PopescuHN}, it did not show teleportation activation as the initial state (introduced in~\cite{Gisin}) is already useful for teleportation before filtering. Generally, a better understanding of the connection between hidden nonlocality and HTP (see also~\cite{HorodeckiTB,Cavalcanti:PRA:2013}) is surely welcome. And what if we allow local filtering on multiple copies of the same state? For Bell-nonlocality~\cite{Brunner:RMP:2014}, this is known~\cite{Peres:PRA:1996} to be useful but its effectiveness for the teleportation-power-activation problem remains to be clarified (see, however,~\cite{Masanes}). To conclude, the possibility of boosting teleportation power beyond the classical threshold is a manifestation of the {\em usefulness} of the {\em shared entanglement}, not only for the task of teleportation but presumably also for other tasks that rely on teleportation as a primitive.

\begin{acknowledgments}
We are grateful to Antonio Ac\'in, Jebarathinam Chellasamy, Kai Chen, Yu-Ao Chen, Huan-Yu Ku, and Marco T\'ulio Quintino, for stimulating discussions and to three anonymous referees for providing very useful comments on an earlier version of this manuscript. This work is supported by the Ministry of Science and Technology, Taiwan (Grants No. 107-2112-M-006-005-MY2, 107-2627-E-006-001, 108-2627-E-006-001, 109-2627-M-006-004, and 109-2112-M-006-010-MY3). H. L., X.-X. F. and T. Z. were supported by the National Natural Science Foundation of China (No. 11974213), National Key R \& D Program of China (No. 2019YFA0308200) and Shandong Provincial Natural Science Foundation (No. ZR2019MA001), and Major Program of Shandong Province Natural Science Foundation (No. ZR2018ZB0649). H. L. was partial supported by Major Program of Shandong Province Natural Science Foundation (Grant No. ZR2018ZB0649).
\end{acknowledgments}

\appendix



\section{Detailed results for Werner states}
\label{App:Werner}

For ease of reference, we reproduce here the FEF of Werner state $\rhoW$ derived in~\cite{Zhao_2010}:
\begin{equation}
  F_{d}[\rhoW] =\left\{
    \begin{array}{cl}
      \tfrac{2v}{d(d+1)},     &  \tfrac{d+1}{2d} \le v \le1,\\[1em]
      \tfrac{2(1-v)}{d(d-1)}, &  0               \le v \le\tfrac{d+1}{2d}, \quad d \text{ even},\\[1 em]
      \tfrac{2 d (1-v)+2}{d^2 (d+1)}, & 0\le v\le\tfrac{d+1}{2d}, \quad d \text{ odd},
    \end{array}\right.
\end{equation}

For the optimizing qubit filter that we have found, it can be shown that the success probability of filtering is
\begin{equation}
	\pW=\tfrac{2[(d+1)(1-v)+3v(d-1)]}{d(d^2-1)}
\end{equation}
while the corresponding increase in FEF for  $0\le v\le \frac{d+1}{4d-2} $ is:
\begin{align}
\nonumber
  F_d &[\rhoWf]-F_d[\rhoW] \\
  &=
    \begin{cases}
    \frac{(d\pW-2)(d^2\pW+dp-6)}{2(d-2)d^2\pW} & \text{(for even $d$)}\\[1em]
      \frac{12-2(d^{2}+4d-4)\pW+(d-1)d^{2}\pW^{2}}{2(d-2)d^{2}\pW} &  \text{(for odd $d$)}.
    \end{cases}
\end{align}
For $v\in\left[\frac{d+1}{4d-2},\frac{1}{2}\right]$, our \ filter could not result in an entangled $\rhoWf$ that beats the classical threshold $F_c$ (they do not seem to exhibit teleportation power).

\begin{figure*}[h!tbp]
\includegraphics[scale=0.39]{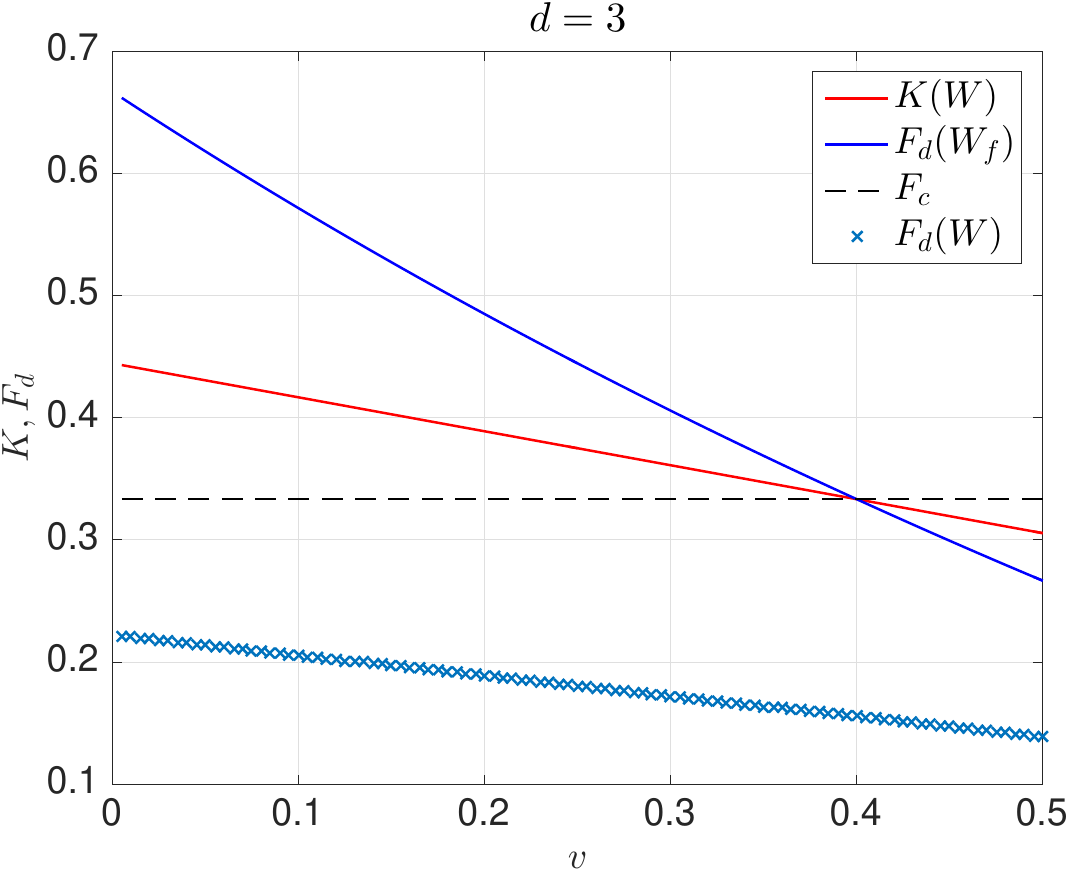}
\includegraphics[scale=0.39]{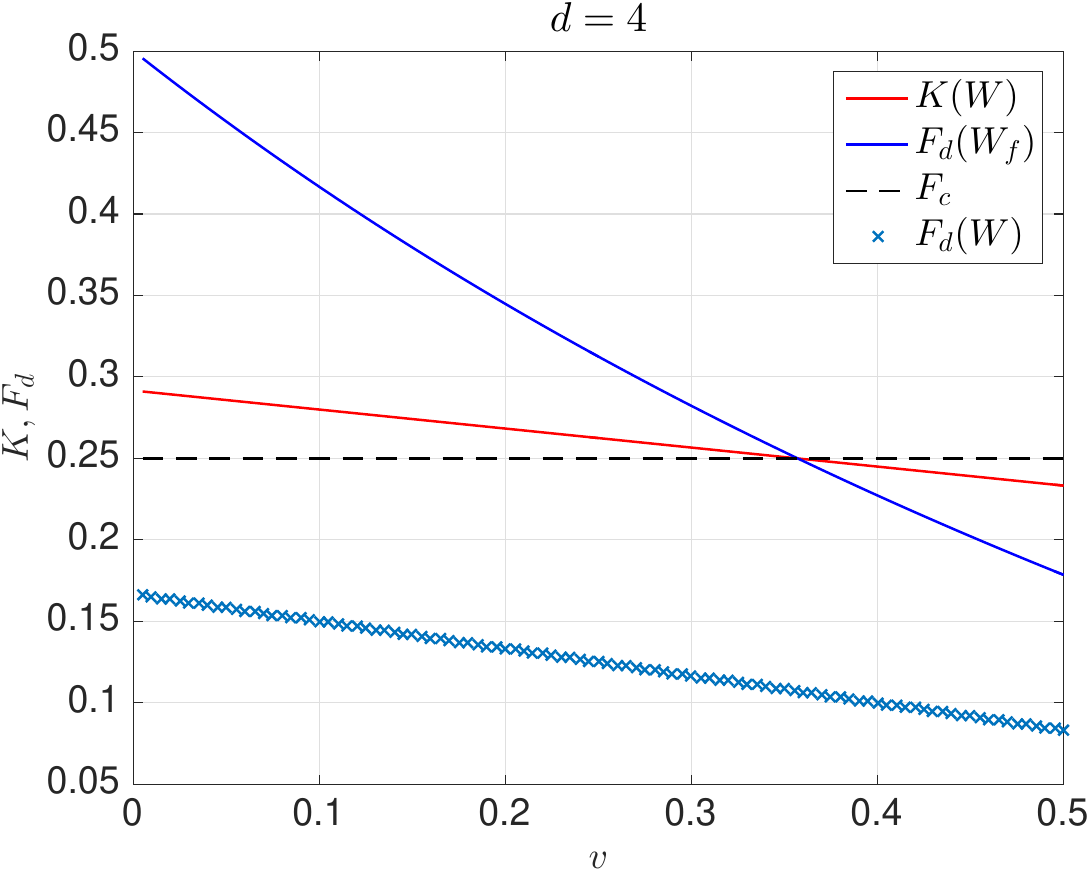}
\includegraphics[scale=0.39]{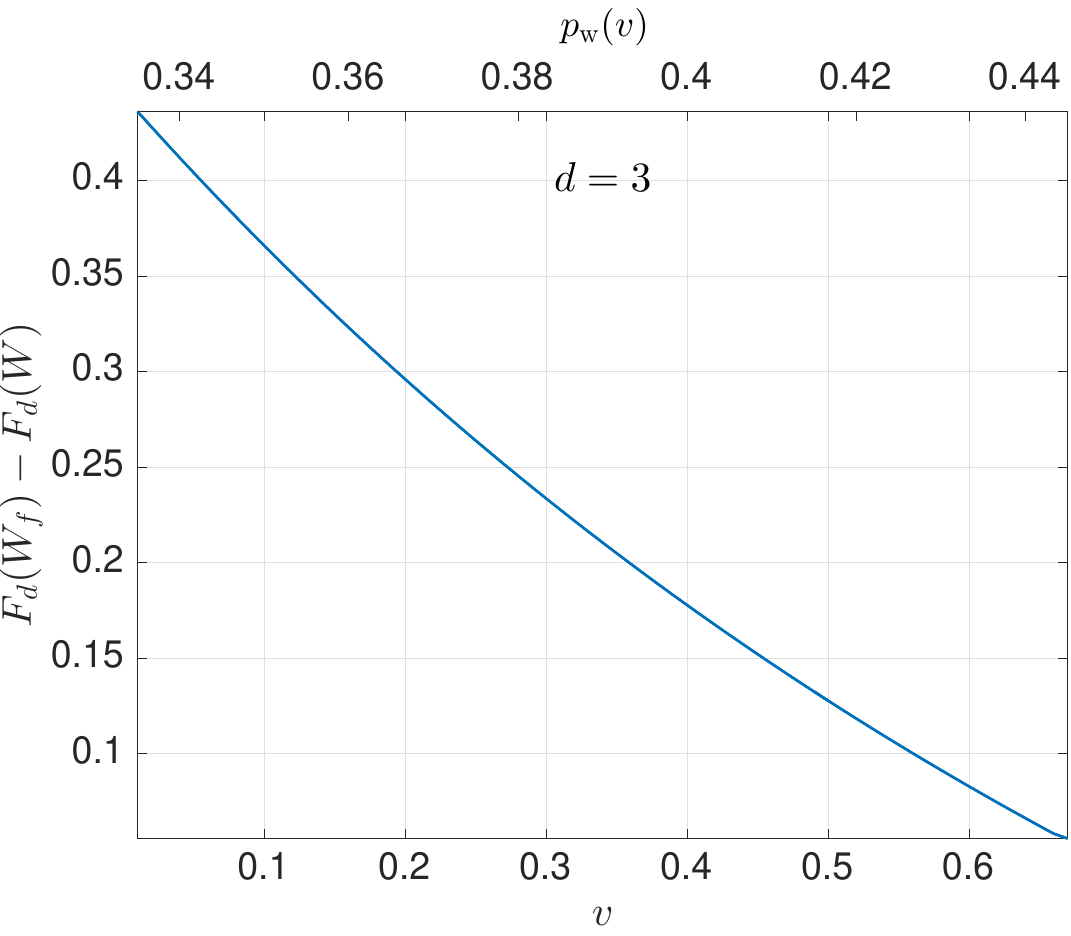}
\includegraphics[scale=0.39]{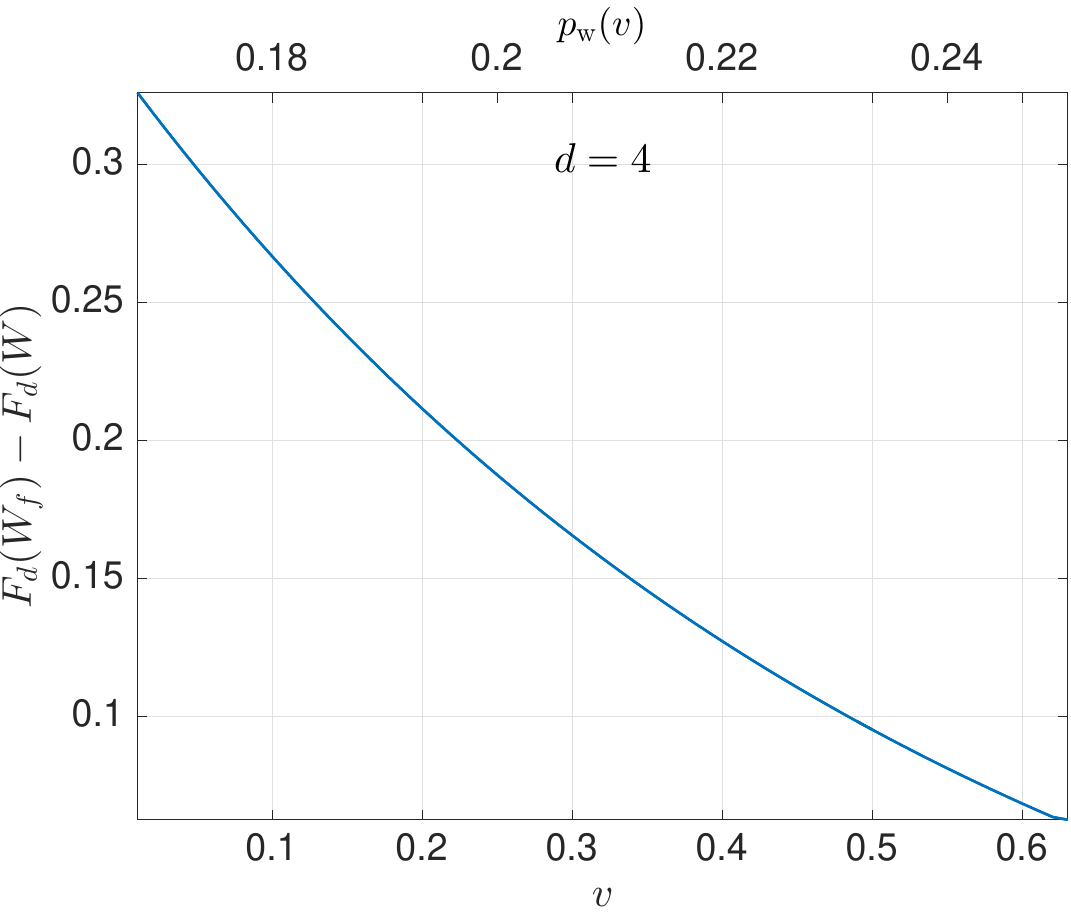}
\caption{(Top) FEF of the Werner state before filtering (blue markers) and after filtering (red solid line for the cost function and blue dash-dotted line for the filtered state). (Bottom) Change in FEF as a function of $v$, which depends linearly on the probability of success in filtering $\pW$.
}
\label{Fig:Werner}
\end{figure*}

In ~\cref{Fig:Werner}, we show, for $d=3$ and $d=4$, the FEF of Werner states before and after filtering, as well as the corresponding cost function. In the same figure, we also show the  difference in FEF, i.e., $F_d[\rhoWf]-F_d[\rhoW]$ vs $v$ [and hence $\pW$, which depends linearly on $v$]. Clearly, when the success probability $\pW$ increases, the amount of FEF that can be increased by local filtering decreases, thus exhibiting some kind of trade-off between these two quantities. The respective plots for larger values of $d$ look similar and are thus omitted.


\section{Detailed results for rank-deficient states}
\label{App:RankTwo}

\subsection{Fully-entangled fraction of $\rho(q)$}
\label{App:FEF:Rhoq}

Here we show that the family of rank-deficient states
in~\cref{Eq:rhoq} is already useful for teleportation whenever (1) $d\ge 4$, or (2)  $d\le 3$ and $q\in\left(\frac{1}{d},1\right)$.

\begin{proof}
Determining $F_d[\rhoq]$ requires the maximization of
\begin{equation}\label{Eq:Overlap}
	\bra{\Psi_d}\rhoq\ket{\Psi_d}=q|\bracket{\Psi_d}{\Phi^+_d}|^2+(1-q)|\bracket{\Psi_d}{01}|^2
\end{equation}
over unitary matrix $U$ such that $\ket{\Psi_d} = (\id\otimes U)\mes$. From \cref{Eq:Overlap} and the form of $\ket{\Psi_d}$, any $U$ that maps $\ket{0},\ket{1}$ outside $S_{01} =\mathrm{span}\left\{ \ket{0},\ket{1} \right\}$ would be suboptimal, since it decreases---when compared with one that acts only nontrivially in $S_{01}$--- the overlap $|\bracket{\Psi_d}{\Phi^+_d}|^2$ and $|\bracket{\Psi_d}{01}|^2$.

Consequently, let us consider only $U$ of the form
\begin{align}\label{eq.maxentstaterhoq}
	U &=
	\begin{pmatrix}
	a & -\bar{b} \\
	b & \bar{a}
	\end{pmatrix}
	\oplus \id_{d-2},
\end{align}
where $a,b\in\mathbb{C}$, $\bar{a}$ ($\bar{b}$) denotes complex conjugation of $a$ ($b$), and the unitary requirement implies that $|a|^2 + |b|^2 = 1$.  Evaluating the overlap gives
\begin{align}
\label{eq.psioverlap}
\nonumber
d\bra{\Psi_d}\rho(q)\ket{\Psi_d} &=   \frac{q}{d} \left[ 4|a|^2  +4(d-2)\mathrm{Re}[a]
   + (d-2)^2 \right] \\ &\quad + (1-q)|b|^2.
\end{align}
Since $d\ge 2$, in maximizing this overlap, we may without loss of generality consider real-valued $a$ and real-valued $b$.
For convenience, let us define
\begin{equation}\label{fofdqa}
  f(d,q,a) :=  \frac{q}{d^2} \left[ 2a + (d-2) \right]^2  + \frac{(1-q)(1-a^2)}{d}.
\end{equation}
Then, we have $F_d[\rhoq] = \max_a f(d,q,a)$.

Using standard variational technique, we find that the local extremum of $f(d,q,a)$ occurs at $a^* = \frac{2(d-2)q}{d(1-q)-4q}$. Note that $|a^*|\le 1$ \Iff  $q$ lies in the interval $\Q:=(0,\frac{1}{3}]\cup[q_0,1]$ where $q_0=\frac{d}{8-d}$. Evaluating $f(d,q,a)$ for $a=a^*$ and
the boundary points $a=0,1$ gives
\begin{subequations}\label{Eq:f}
\begin{align}
	\label{Eq:F-a*}
	f(d,q,a^*) &= \frac{(1-q)[(d-5)q+1]}{d(1-q)-4q},\quad q\in\Q,\\
	\label{Eq:F-a0}
	f(d,q,0) &= \frac{d^2q -5dq+d+4q}{d^2}, \\
	\label{Eq:F-a1}
	f(d,q,1) &= q.
\end{align}
\end{subequations}
Taking their difference gives
\begin{subequations}\label{Eq:Diff}
\begin{align}
	\label{Eq:01}
	f(d,q,1)-f(d,q,0) &= \frac{q(5d-4)-d}{d^2}, \\
	\label{Eq:astar-0}
	f(d,q,a^*)-f(d,q,0) &= \frac{4(d-2)^2 q^2}{d^2[d(1-q)-4q]}, \\
	\label{Eq:astar-1}
	f(d,q,a^*)-f(d,q,1) &= \frac{(1-3q)^2}{d(1-q)-4q},
\end{align}
\end{subequations}
where we note that the last two equations are only meaningful for $q\in\Q$.

For $d=2$, \cref{Eq:astar-0} vanishes and~\cref{Eq:01} is non-positive \Iff  $q\in(0,\frac{1}{3}]$. For $d= 3$, \cref{Eq:astar-0} is positive for $q\in(0,\frac{1}{3}]$ while ~\cref{Eq:F-a1} dominates for other values of $q\in(0,1]$.

For $d\ge 4$, $\Q=(0,\frac{1}{3}]$ since $|q_0|\ge 1$. Then, for $q\in\Q$, one has $d(1-q)-4q>0$ and thus $f(d,q,a^*$) dominates over the other expressions in \cref{Eq:f}. For the complementary interval $q\in(\frac{1}{3},1]$,  \cref{Eq:01} is positive and $f(d,q,1)$ dominates in this interval. Putting everything together, we thus have
\begin{equation}\label{Eq:FdRhoq}
	F_d[\rhoq] =
	\begin{cases}
	\frac{(1-q)[(d-5)q+1]}{d(1-q)-4q}, & 0 \le q \le \frac{1}{3}, \\
	q, & q > \frac{1}{3}.
	\end{cases}
\end{equation}

To determine the dimension $d$ for which $\rho(q)$ is always useful for teleportation, it is expedient to consider the function
\begin{equation}
	G(d,q) = d F_d[\rho(q)] - 1 = \frac{q[d(d-5)(1-q) + 4]}{d(1-q)-4q},
\end{equation}
where the last equality holds for $0 \le q \le \frac{1}{3}$. For the complementary interval of $q>\frac{1}{3}$, it is straightforward to determine when $F_d[\rhoq]>\frac{1}{d}$ and hence useful for teleportation. Coming back to $q\in[0,\frac{1}{3}]$, we see that $F_d[\rhoq] > \od$ \Iff  $G(d,q) > 0$. When $d \geq 5$, $G(d,q)>0$ since both numerator and denominator are positive for $0 < q \le \od$.  Similarly, for $d=4$, $G(d,q)$ simplifies to $\frac{q^2}{1-2q}$, which is strictly positive for $0 < q \le \od=\frac{1}{4}<\frac{1}{2}$. Hence, as claimed, $F_d[\rhoq]>\frac{1}{d}$ for $d\ge 4$ and $q\in(0,1]$, i.e., these states are all useful for teleportation even before  filtering.

For the case of $d=3$, we have $G(d,q) = \frac{2q(1-3q)}{7q-3}$, which is easily verified to be {\em non}-positive for $q\in(0,\od)$. Together with \cref{Eq:FdRhoq}, we thus see that $\rhoq$ for $d=3$ is useless for teleportation \Iff  $q\in(0,\frac{1}{3}]$. Finally, $G(2,q)=-q<0$ and thus $\rhoq$ for $d=2$ is useless for teleportation \Iff  $q\in(0,\frac{1}{2}]$.

\end{proof}

\subsection{Violating the reduction criterion}
\label{App:VReduction}

A bipartite state $\rho_{AB}$ acting on $\mathbb{C}^d\otimes\mathbb{C}^d$  satisfies the reduction criterion of separability (RC)  if
\begin{align}\label{Eq:RC}
    \rho_A \otimes \id_d - \rho_{AB} \succeq 0  \quad\text{ and }\quad
    \id_d \otimes \rho_B - \rho_{AB} \succeq 0,
\end{align}
where $\rho_A$ and $\rho_B$ are, respectively, the reduced density matrix on Alice's and Bob's side while $\succeq0$ means matrix positivity. Here we show that for all $q \in (0,1]$,
the rank-deficient states  violate the first condition.

With $\rho_{AB} = \rho(q)$ then
$\rho_{A} = q \frac{\id_d}{d} + (1-q) \proj{0}$.
Let
\begin{align}
\nonumber
    R &= \rho_{A}\otimes \id_d - \rho_{AB} \\
    &= \frac{q}{d}
    \left( \id_d\otimes \id_d - d \proj{\Psi_{d}} \right)
    + (1-q) \sum_{j \ne 1} \proj{0j}
\end{align}

Note we can decompose $R$ as the sum of two Hermitian matrices, i.e., $R = R_s + R_d$ where
\begin{align}
    R_s &=  \frac{q}{d}
    \left(  \id_d\otimes  \id_d - d \proj{\Psi_{d}} \right)
    + (1-q) \proj{00},
\end{align}
and $R_d = (1-q) \sum_{j \ne 0,1} \proj{0j}$.
We next show that one can find an eigenvector of $R$ in the subspace  $\mathcal{S} = \mathrm{span}\{ \ket{jj}: j = 0,1,..,d-1 \}$ with negative eigenvalue. Note that $R_d\mathcal{S}=0$, so it suffices to restrict our attention to  $R_s$ in the following discussion.

In the subspace $\mathcal{S}$, $R_s$ can be represented in the basis $\ket{jj}$ as a sum of a diagonal matrix $D$ and constant matrix $C$:
\begin{align}
    D &\hat{=} \mathrm{diag}\left(1-q + \frac{q}{d}, \frac{q}{d}, ..., \frac{q}{d} \right), &
    C &\hat{=} -\frac{q}{d} J,
\end{align}
where $J$ is a $d \times d$ all-ones matrix. That is, in the subspace $\mathcal{S}$, the matrix $R_s$ has zeros on the diagonal except the $\proj{00}$ component, and $(-\frac{q}{d})$ on all off-diagonal terms.

Consider the (un-normalized) vector
\begin{equation}
    \ket{\psi} = t \ket{00} + \sum_{j} \ket{jj}
    \hat{=} (t, 1, \ldots, 1)^T.
\end{equation}
Let $\beta = -\frac{q}{d}$.
From the eigenvalue equation
$R_{s} \ket{\psi} = \lambda \ket{\psi}$, we have
\begin{align}
    (1-q) t + \beta(d-1) &= \lambda t, &
    \beta t+\beta(d-2) &= \lambda.
\end{align}
Eliminating $t$, we obtain
\begin{equation}
    [\lambda - (1-q)][\lambda - \beta(d-2)] - \beta^2 (d-1) = 0
\end{equation}
This is a quadratic equation $\lambda^2 + b\lambda + c = 0$
with
\begin{align}
\nonumber
    b &= - [(1-q) + \beta(d-2) ], \\
    c &= (1-q)\beta(d-2) -\beta^2(d-1).
\end{align}
We have that $(1-q) \ge 0 $ and $\beta < 0$ for $q \in (0, 1]$.
It can be checked that the discriminant $\Delta = b^2 - 4c$ is
\begin{align}
    \Delta &= (1-q)^2 + 4\beta(1-q)
    -2d\beta (1-q) + d\beta^2 > 0
\end{align}
when $q \in (0, 1]$ so we have two distinct real roots.

For $d \ge 2$ and $q \in (0, 1]$ we see that $c < 0$. But this is the product of the two roots so they must have opposite sign.  Thus $R_{s}$, and hence $R$ has an eigenvector $\ket{\psi}\in\mathcal{S}$ with negative eigenvalue $\lambda$.  In other words, for $\rho(q)$, the left-hand-side of the first inequality is violated, i.e., it violates RC.

\subsection{Optimal one-side filter for maximizing the cost-function}
\label{App:OptimalOneSide}

Here we prove that if we restrict to one-side local filtering, then \cref{Eq:OptimalFilterRhoq} is optimal for maximizing the cost-function of
$\rho(q)$ with local dimension $d$. Let the unnormalized filtered state and the probability of success in filtering be
\begin{align}
	\tau(q) &= (A\otimes \id_d) \rho(q) (A \otimes \id_d)^\dag, &
	p(q) &= \tr[\tau(q)].
\end{align}
Recall that the cost-function may be written as
\begin{equation}
	K_d(q) := K[\rho(q)] = \bra{\Phi_d^+} \tau(q) \ket{\Phi_d^+} + \frac{1-p(q)}{d}.
\end{equation}
Our goal is to find the one-side filter $A$ that maximizes $K_d(q)$ under the constraint that $\lVert{A}\rVert_\infty = 1$ (i.e., the maximum singular value of $A$ is 1).

 Let $M = A^\dag A$ and $m_i$ be the nonzero eigenvalues of $M$.  Suppose for now that $M$ is upper triangular, then
\begin{equation}
m_i = M_{ii} = \sum_{j} (A^\dag)_{ij} A_{ji} = \sum_{j} A^*_{ji}A_{ji} = \sum_{j} \lvert A_{ji}\rvert^2.
\end{equation}
Note that the non-zero singular values of $A$ are given by the positive square roots of the non-zero eigenvalues of $A^\dag A$.
Thus, the constraint $\lVert{A}\rVert_\infty = 1$ amounts to requiring
\begin{equation}\label{Ineq:Constraint}
\max_{j} m_j = 1  \, \implies \,
\sum_{i} \lvert A_{ij} \rvert^2 \le 1\,\,\forall\,\,j.
\end{equation}

If $M$ is not upper triangular, then from Schur decomposition, one can always find some
unitary $Q$ such that
\begin{equation}
\tilde{M} = Q M Q^\dag
\end{equation}
is upper triangular.
Note $M$ and $\tilde{M}$ have the same eigenvalues since they are unitarily related.
Also we have that
\begin{equation}
\tilde{M} = Q(A^\dag A )Q^\dag = (QA^\dag Q^\dag) (Q A Q^\dag) = \tilde{A}^\dag  \tilde{A}
\end{equation}
so the same unitary $Q$ relates $A$ and $\tilde{A}$. Hence, the implication of~\cref{Ineq:Constraint} holds for a general filter $A$.

In terms of the filter matrix elements $A_{ij} \in \mathbb{C}$,
\begin{equation}
\begin{split}\label{Eq:Costfunction}
  K_d(q) &= \frac{1}{d} + \frac{q}{d^2}\left| \sum_i A_{ii} \right|^2
        - \frac{q}{d^2} \sum_{ij} \left| A_{ij} \right|^2  \\
    &\quad - \frac{(1-q)}{d}\sum_j |A_{j1}|^2
        + \frac{(1-q)}{d}|A_{21}|^2.
\end{split}
\end{equation}
Note that the contribution of each off-diagonal term is always negative. Moreover, the constraint of~\cref{Ineq:Constraint} puts a limit on the sum $|A_{ij}|^2$ for matrix elements in the same column $j$, i.e., one can only increase the magnitude of the diagonal entries $|A_{jj}|$ by reducing  the magnitude of off-diagonal elements $|A_{ij}|, i \ne j$ in the same column.

Hence, the optimal one-side local filter must be diagonal. This allows us to simplify the cost function, via~\cref{Eq:Costfunction} to:
\begin{align}
\label{eq.offdiag}
\tilde{K}_d(q) &= \ K_d(q) \text{ with all $A_{ij}=0$ for $i\neq j$}\\
\nonumber
 &=  \frac{1}{d} + \frac{q}{d^2}\left| \sum_i A_{ii} \right|^2
    - \frac{q}{d^2} \sum_i \left| A_{ii} \right|^2 - \frac{(1-q)}{d}|A_{11}|^2.
\end{align}
Let $A_{ii} = x_i + {\rm i}\, y_i$, where $x_i,y_i \in \mathbb{R}$ for $i = 1,2,\ldots, d$.
Then we have
\begin{align}\label{eq.diag}
\nonumber
\tilde{K}_d(q) &= \frac{1}{d} + \frac{q}{d^2} \sum_{i \neq j}x_{i} x_{j} - \frac{(1-q)}{d}x_{1}^2 \\
 &\quad     + \frac{q}{d^2} \sum_{i \neq j}y_{i} y_{j} - \frac{(1-q)}{d}y_{1}^2
\end{align}

Clearly, $\tilde{K}_d(q)$ is linear in both $x_i$ and $y_i$ for all $i>1$. Moreover, the constraint $\lVert{A}\rVert_\infty = 1$ for a diagonal $A$ means that we must have $x_i^2 + y_i^2 \le 1$ for all $i$. Then, from the form of $\tilde{K}_d(q)$ and the fact that $0\le x_i,y_i\le 1$, it is clear that the maximization of $\tilde{K}_d(q)$ can be attained by setting $x_i=1$ and $y_1=y_i=0$ for all $i>2$, thereby giving 
\begin{equation}
\tilde{K}_d(q) = \frac{1}{d} + \frac{q}{d^2}\left[(d-1)^2+ 2x_1(d-1)\right] - \frac{(1-q)}{d}x_{1}^2
\end{equation}

Then, standard variational arguments imply that a one-side filter maximizing $K_d(q)$ is diagonal, taking the form of
\begin{align}
A=\text{diag}[\tfrac{(d-1)q}{d(1-q)}, 1,\cdots,1]\quad \text{for } q\in(0,\tfrac{d}{2d-1}),
\end{align}
whereas the optimal filter is the identity operator for $q\in[\frac{d}{2d-1},1]$.

\subsection{Two-side filtering (quasidistillation)}
\label{App:TwoSide}

In~\cite{Horodecki}, the family of local filters $A_n=\mathrm{diag}[1/n,1,...,1]$,  $B_n=\mathrm{diag}[1,1/n,...,1/n]$ were proposed to quasi-distill $\rho(q)$ into $\mes$. From some simple calculation, one finds that these filters yield the unnormalized state\footnote{Note that it was claimed in Eq.~(40) in~\cite{Horodecki} that the filtered state takes the form of $\frac{1}{n}\left[q\proj{\Phi_d^+}+\left(\frac{1-q}{n}\right)\proj{0}\otimes\proj{1} \right]$, which is incorrect.}
\begin{equation}
	\tau_n=\frac{1}{n^2} \left[ q\proj{\Phi_d^+}+\left(\frac{1-q}{n^2}\right)\proj{0}\otimes\proj{1} \right]
\end{equation}
with a success probability of $p_n=\frac{q(n^2-1)+1}{n^4}$. For $n\gg1$, the FEF is attained by taking the overlap with $\mes$, then
\begin{equation}
    F_d \left(\frac{\tau_n}{p_n}\right) = \frac{q}{n^2p_n} = 1 - \frac{1-q}{q(n^2-1) + 1}.
\end{equation}
Thus, when $n\to\infty$, $F_d \left(\frac{\tau_n}{p_n}\right)\to 1$ but the success probability $\lim_{n\to\infty} p_n= 0$.

\begin{figure*}[h!tbp]
 \includegraphics[scale=0.35]{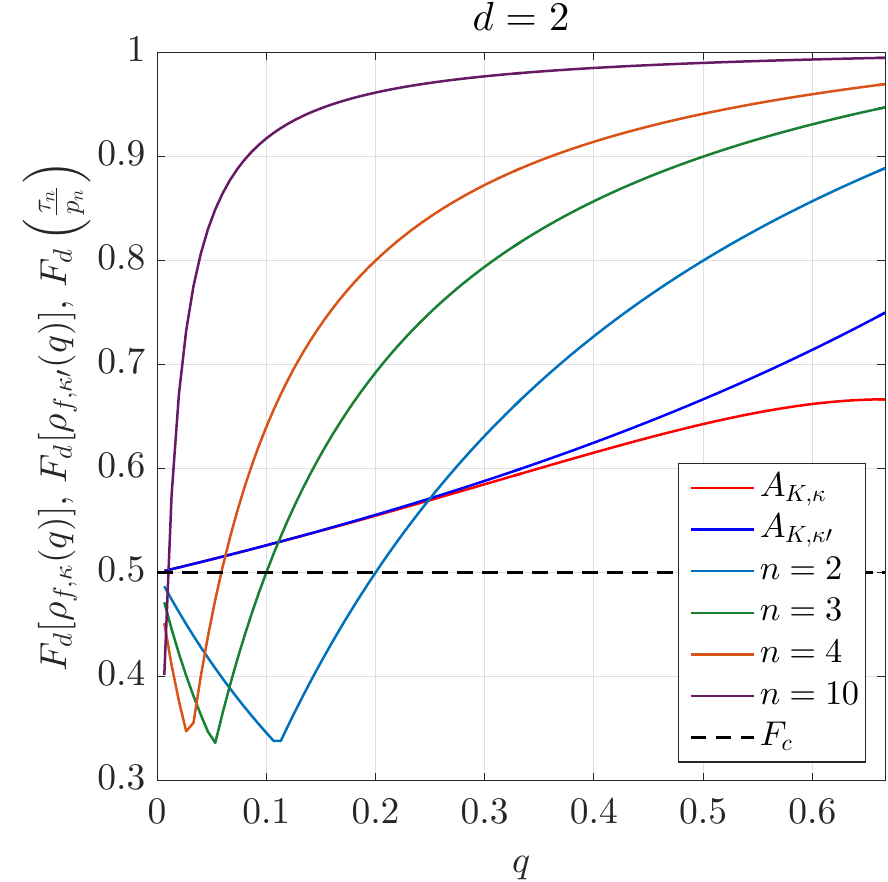} \hspace{0.05cm}
 \includegraphics[scale=0.35]{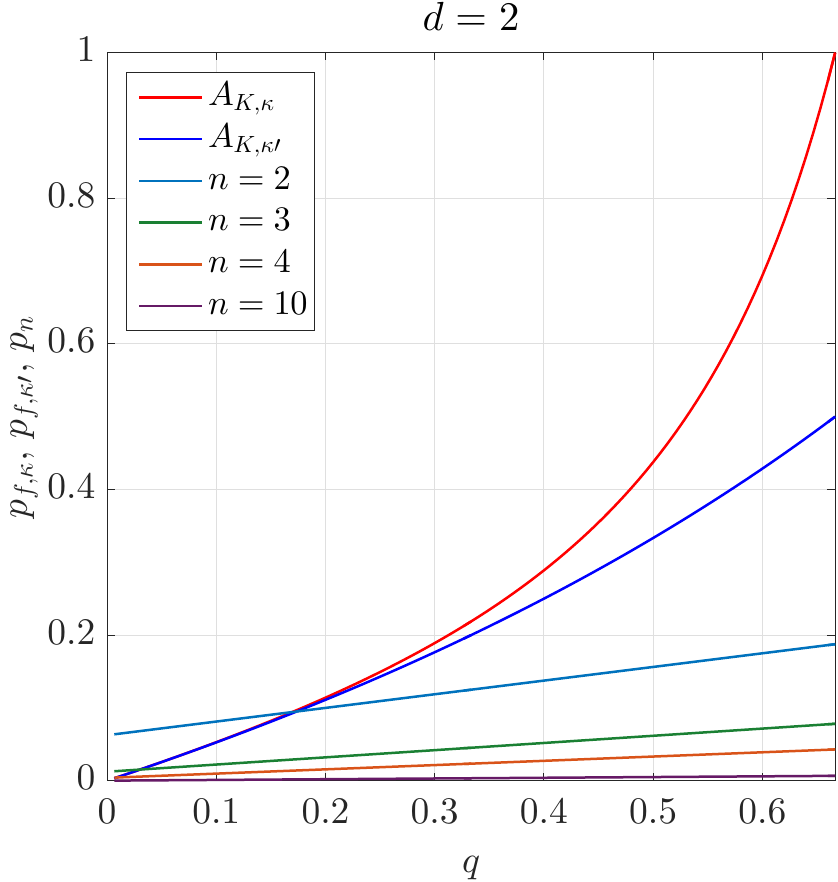} \hspace{0.05cm}
 \includegraphics[scale=0.35]{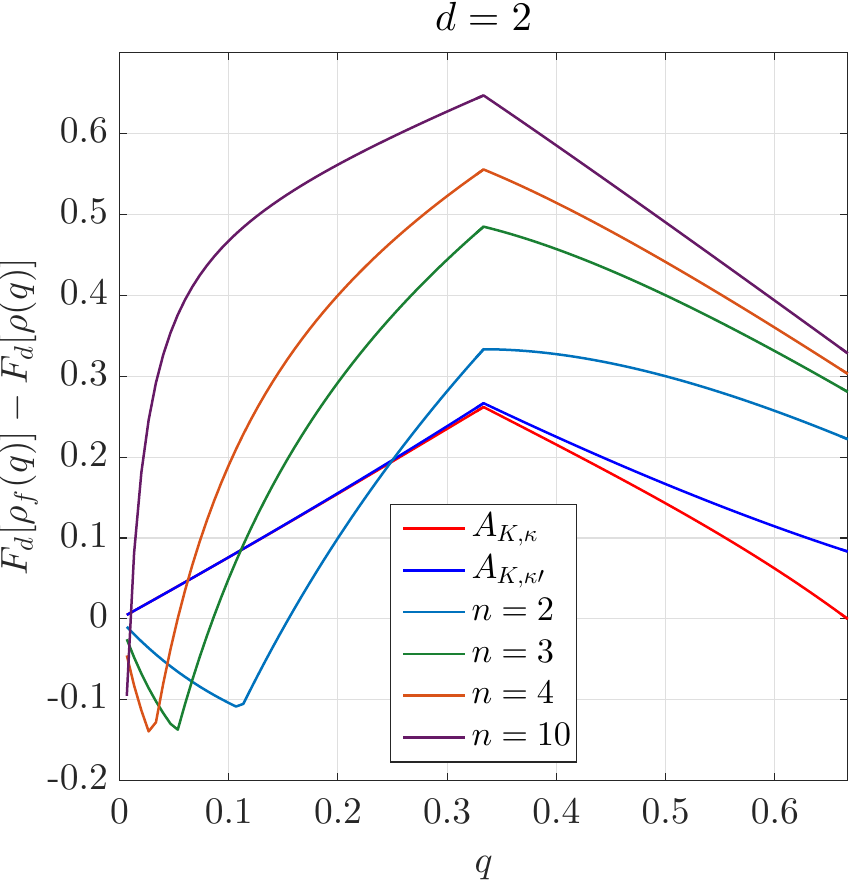} \\
\includegraphics[scale=0.35]{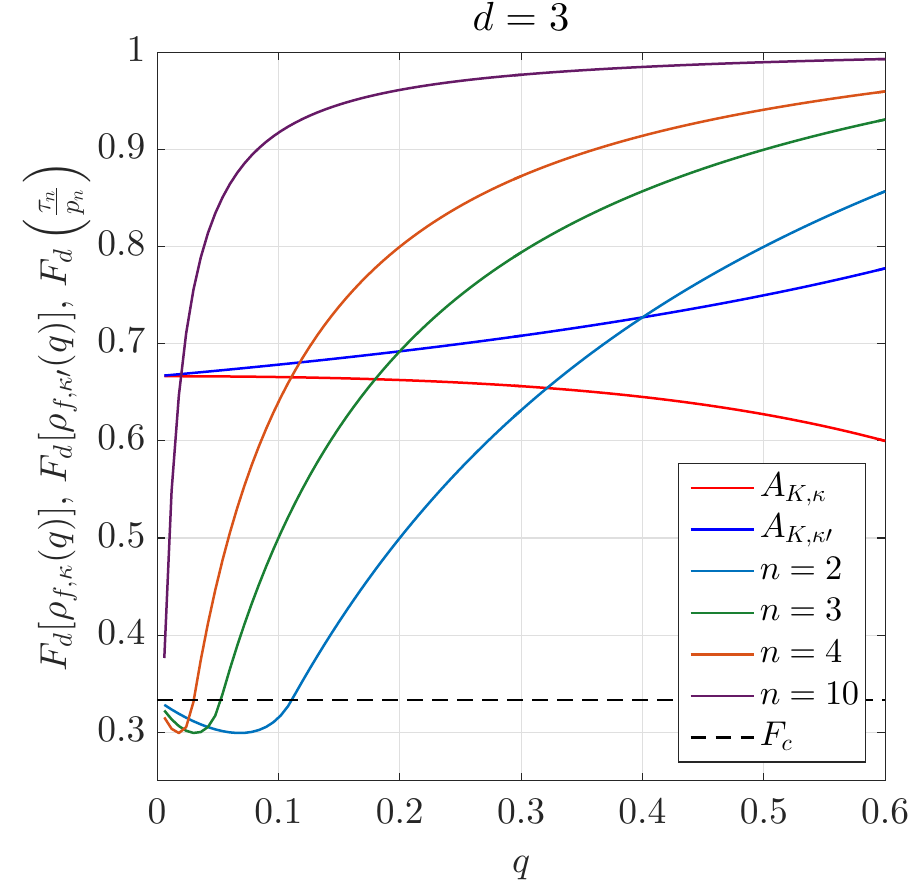} \hspace{0.05cm}
 \includegraphics[scale=0.35]{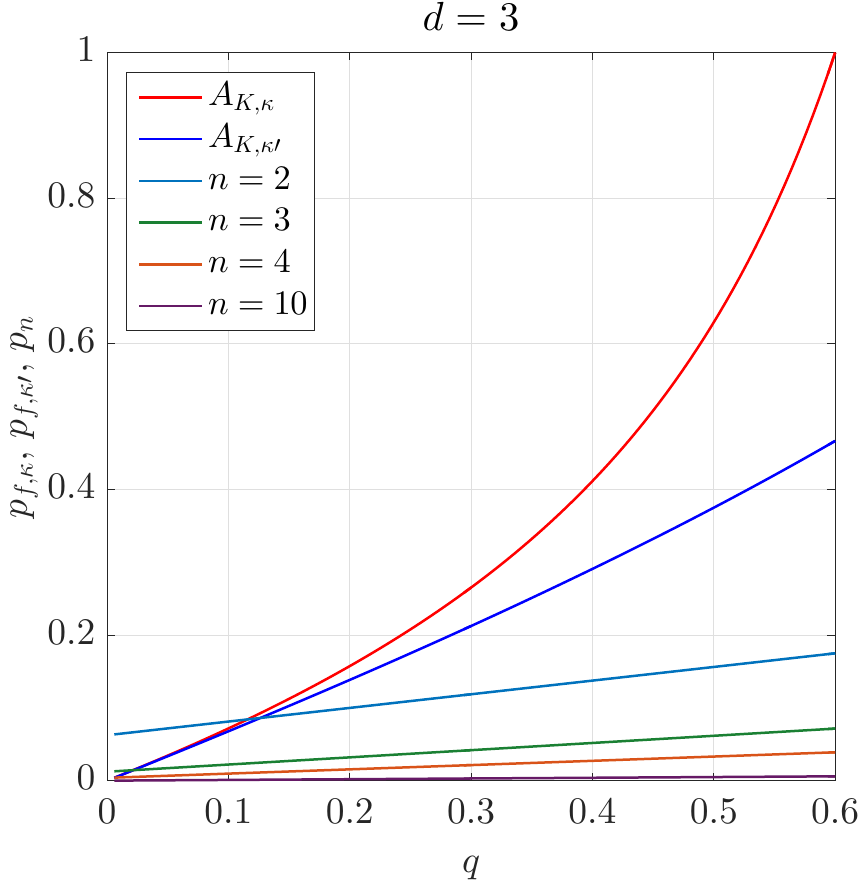}  \hspace{0.05cm}
 \includegraphics[scale=0.35]{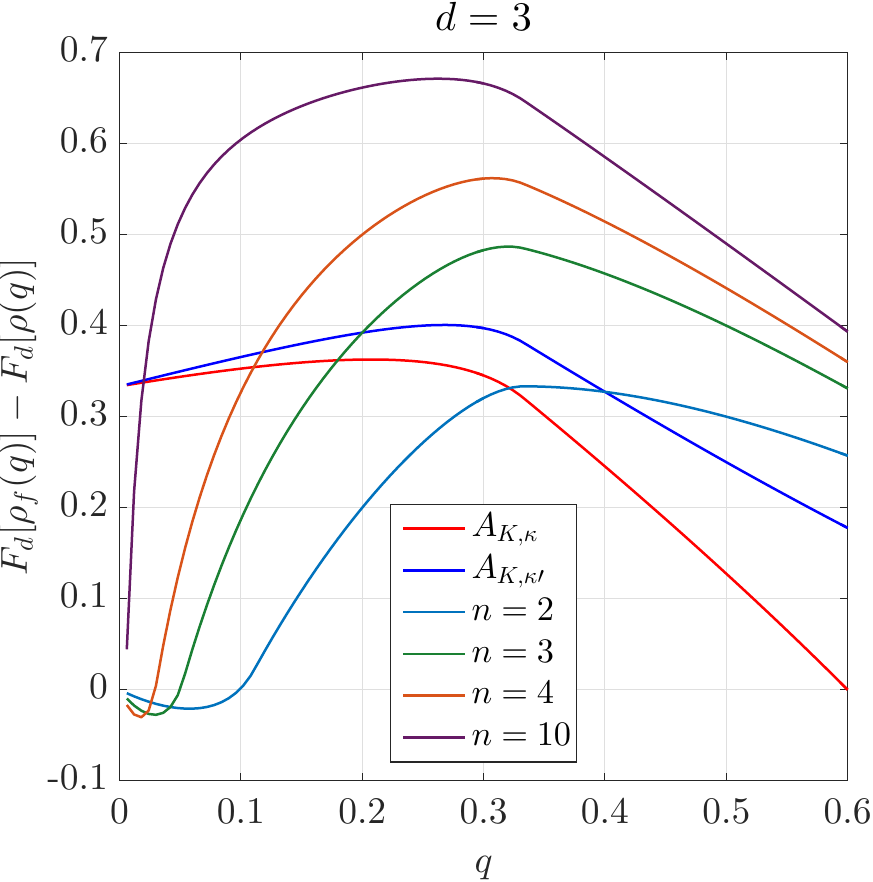}
 \caption{(Left) Comparison between the FEF of the filtered state $\rhof$ obtained by employing different filtering schemes on $\rhoq$ for $d=2$ and $d=3$. Included in the plots are the two-side filtering schemes introduced in~\cite{Horodecki} for $n=2,3,5,$ and 10 (see~\cref{App:TwoSide}) as well as the single-side filtering schemes discussed in \cref{App:OptimalOneSide}. (Center) Comparison of the corresponding success probabilities as a function of the parameter $q$. (Right) Comparison of the corresponding change in FEF as a function of the parameter $q$. }
 \label{Fig:rhoq-fidelity-compare}
 \end{figure*}

For the performance of these filters against the one-side filters discussed in~\cref{App:OptimalOneSide}, see~\cref{Fig:rhoq-fidelity-compare}.

\section{Experimental details}
\label{App:Exp}

In this Appendix, we provide further details about our experimental setup. A schematic, simplified version of this setup that emphasizes its connection with the teleportation protocol can be found in Fig. 2a whereas an overview of the full experimental setup is given in Fig.~2b. In the following subsections, we explain how each of the boxed section in Fig.~2b functions. To this end, it would be useful to bear in mind the following:

\begin{enumerate}
\item[(i)] A half-wave plate (HWP) @ $\theta$ performs the unitary transformation $U_\text{\tiny HWP}=\cos2\theta(\proj{H}-\proj{V})+\sin2\theta(\ket{H}\!\bra{V}+\ket{V}\!\bra{H})$ on a polarization state, where $\theta$ is the angle between the fast axis of the HWP and the vertical direction. \vspace{-0.2cm}
\item[(ii)] A beam displacer (BD) transmits a vertically polarized photon but deviates a horizontally polarized one.
\item[(iii)] A polarizing beam splitter (PBS) transmits a horizontally polarized photon but reflects a vertically polarized one.
\item[(iv)] A quarter-wave plate (QWP) @ $\theta$ performs the unitary transformation $U_\text{\tiny QWP}=\frac{1}{\sqrt{2}}[\id_2+{\rm i }\cos2\theta(\proj{H}-\proj{V})+{\rm i }\sin2\theta(\ket{H}\!\bra{V}+\ket{V}\!\bra{H})]$, on a polarization state where $\id_2=\proj{H}+\proj{V}$ and $\theta$ is the angle between fast axis of the QWP and the vertical direction.
\end{enumerate}

\subsection{Entangled photon source}

\begin{figure}[h!tbp]
\includegraphics[width=0.82\columnwidth]{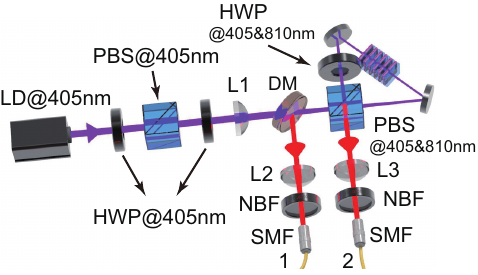}\hspace{1cm}
\includegraphics[width=0.82\columnwidth]{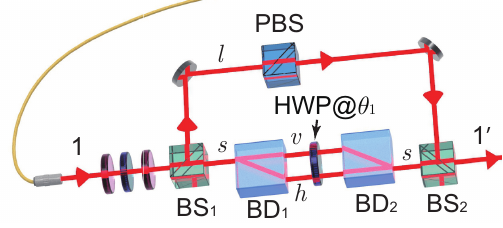}
\caption{Zoom-in view of the (top) ``Photon source" part and the (bottom) ``Noisy channel" part of Fig. 2b. The top setup aims to generate photon pairs maximally entangled in the polarization degree of freedom (DOF) whereas the bottom setup aims to generate, starting from photon pairs produced using the first setup, two-qubit mixed quantum states $\rhoq$ [see \cref{Eq:rhoq}] encoded in the polarization DOF.}
\label{Fig:PhotonSource}
\end{figure}

We start by describing how polarization-entangled photon pairs are produced in our setup by bidirectionally pumping a periodically poled potassium titanyl phosphate (PPKTP) crystal (placed in a Sagnac interferometer~\cite{Kim06}) with an ultraviolet (UV) diode laser at 405~nm. Specifically, as shown in Fig.~\ref{Fig:PhotonSource}, the power of the pump light is first adjusted through a HWP and a PBS. Then, at the second HWP set at 22.5$^\circ$, the horizontal polarization $\ket{H_p}$ is rotated to $\ket{+_p}=\frac{1}{\sqrt{2}}(\ket{H_p}+\ket{V_p})$. Via two lenses L$_{1}$ (with focal length 75~mm and 125~mm), the pump beam is subsequently focused into a beam waist of 74 $\mu$m  and arrives at a dual-wavelength PBS after passing through a dichroic mirror.

The pump beam is then split on the PBS and coherently pumped through the PPKTP in the clockwise and counterclockwise direction.  The PPKTP crystal, with dimensions 10 mm (length) $\times$ 2 mm (width) $\times$ 1 mm (thickness) and a poling period of $\Lambda=10.025~\mu$m, is housed in a copper oven and temperature controlled by a homemade temperature controller set at 29$^{\circ}$C to realize the optimum type-\uppercase\expandafter{\romannumeral2} phase matching at 810~nm. The clockwise and counterclockwise photons are then recombined on the dual-wavelength PBS to generate entangled photons with an ideal form of  $\ket{\Psi^{+}_{12}}=\frac{1}{\sqrt{2}}(\ket{H_1V_2}+\ket{V_1H_2})$.

After that, photon 1 and 2 are filtered by a narrow band filter (NBF) with a full width at half maximum (FWHM) of 3~nm, and coupled into single-mode fiber (SMF) by lenses of focal length 200 mm (L$_2$ and L$_3$) and objective lenses (not shown in Fig.~\ref{Fig:PhotonSource}). During our experiment, the pump power is set at 5~mW, and we observe a two-fold coincidence count rate of 7.3$\times$10$^4$/s.


\subsection{Noisy channel $\mathcal{E}(\theta_1)$}
\label{App:NoisyChannel}

In this part of the experimental setup, which does not involve photon 2 (as can be seen in \cref{Fig:PhotonSource}), photon 1 goes through a noisy channel $\mathcal{E}(\theta_1)$ that eventually results in a  two-photon polarization state given by $\rhoq$ (in the ideal scenario). To this end, photon $1$ is first guided to an unbalanced Mach-Zehnder interferometer (MZI) after passing a polarization controller (PC). Then, BS$_{1}$ transforms an ideal maximally entangled two-qubit state $\ket{\Psi_2^{+}}=\frac{1}{\sqrt{2}}(\ket{H_1V_2}+\ket{V_1H_2})$  to $\frac{1}{2}(\ket{H_1V_2}+\ket{V_1H_2})\otimes(\ket{s_1}+\ket{l_1})$ with $\ket{s_1}$ and $\ket{l_1}$ denote, respectively, the short and long arm of the unbalanced MZI.

On the long arm, the PBS  only transmits $\ket{H_1}$ and filters away the $\ket{V_1}$ component. On the short arm, the two BDs and a HWP (at angle $\theta_1$) work together as an attenuator so that $\ket{s_1}\to\sin^22\theta_1\ket{s_1}$. Indeed, from the property of a BD and the calculation shown in \cref{eq:NoisyChannel}, we see that a photonic state that goes through the short arm is attenuated by a factor of $\sin^22\theta_1$. Since photons that travel through the long arm and those that travel through the short arm are distinguishable, the two spatial modes $\ket{s_1}$ and $\ket{l_1}$ are incoherently recombined at BS$_2$. In the experiment, we keep only photons exiting from the output port 1', thus obtaining the state $\rho_{1^{\prime}2}=q(\theta_1)\ket{\Phi_2^{+}}\bra{\Phi_2^{+}}+(1-q(\theta_1)) \proj{HV}$ with $q(\theta_1)=\frac{2\sin^22\theta_1}{1+2\sin^22\theta_1}$. With this  setup, $q(\theta_1)$ can be tuned in the range from 0 to $\frac{2}{3}$. A step-by-step calculation detailing the evolution of the two-photon state through this setup is given in \cref{eq:NoisyChannel}.

\begin{align}
\ket{\Psi_2^{+}}&=\frac{1}{\sqrt{2}}(\ket{H_1V_2}+\ket{V_1H_2})\nonumber\\
&\xrightarrow{\text{BS}_1}\frac{1}{2}(\ket{H_1V_2}+\ket{V_1H_2})\otimes(\ket{s_1}+\ket{l_1})\nonumber\\
&\xrightarrow[\text{at long arm}]{\text{PBS}}\frac{1}{\sqrt{3}}(\ket{H_1}\ket{V_2}\ket{l_1}+\ket{H_1}\ket{V_2}\ket{s_1}+\ket{V_1}\ket{H_2}\ket{s_1})\nonumber\\
&\xrightarrow[\text{at short arm}]{\text{BD}_1}\frac{1}{\sqrt{3}}(\ket{H_1}\ket{V_2}\ket{l_1}+\ket{H_1}\ket{V_2}\ket{h_1}+\ket{V_1}\ket{H_2}\ket{v_1})\nonumber\\
&\xrightarrow[\text{at short arm}]{\text{HWP}~@~\theta_1}\frac{1}{\sqrt{3}}\big[\ket{H_1}\ket{V_2}\ket{l_1}\nonumber\\
&\qquad\qquad\qquad+(\cos2\theta_1\ket{H_1}+\sin2\theta_1\ket{V_1})\ket{V_2}\ket{h_1}\nonumber\\
&\qquad\qquad\qquad+(\sin2\theta_1\ket{H_1}-\cos2\theta_1\ket{V_1})\ket{H_2}\ket{v_1}\big]\nonumber\\
&\xrightarrow[{+ \text{post-select path }s} \text{ at short arm}]{\text{BD}_2}\frac{1}{\sqrt{1+2\sin^22\theta_1}}(\ket{H_1}\ket{V_2}\ket{l_1}\nonumber\\
&\qquad\qquad\qquad+\sin2\theta_1\ket{H_1}\ket{H_2}\ket{s_1}+\sin2\theta_1\ket{V_1}\ket{V_2}\ket{s_1})\nonumber\\
&\xrightarrow[{\text{incoherently combined}}]{\text{BS}_2}\tfrac{2\sin^22\theta_1}{1+2\sin^22\theta_1}\ket{\Phi_2^{+}}_{1^\prime2}\bra{\Phi_2^{+}}\nonumber\\
&\qquad\qquad\qquad+\tfrac{1}{1+2\sin^22\theta_1}{ \ket{H_{1^\prime}V_2}\bra{H_{1^\prime}V_2}}\label{eq:NoisyChannel}
\end{align}


\subsection{Local filtering}

\begin{figure}[h!tbp]
\includegraphics[width=0.8\columnwidth]{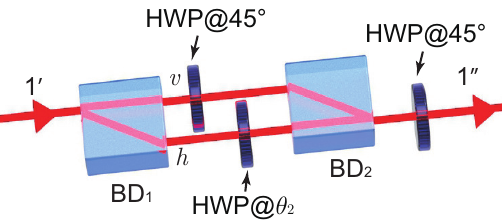}
\caption{Experimental setup (zoom-in view of the ``Filter"  part of Fig.~2b) that performs the filtering operation of \cref{Eq:OptimalFilterRhoq} for the $d=2$ case.}
\label{Fig:LocalFilter}
\end{figure}

Our setup for implementing the local filter $A_\kappa=\text{diag}[\kappa,1]$  is shown in Fig.~\ref{Fig:LocalFilter}. As with the attenuator discussed in \cref{App:NoisyChannel}, this part of the setup consists also of two BDs in addition to three HWPs. For photons encoded in the polarization DOF, filter $A$ attenuates the horizontal component $\ket{H}$ by a factor of $\kappa$ while keeping the vertical component $\ket{V}$ unchanged. To illustrate the effect of this setup, we provide in ~\cref{eq:LocalFilter} a step-by-step calculation showing how a general input polarization {\em pure} state $\alpha\ket{H_{1^{\prime}}}+\beta\ket{V_{1^{\prime}}}$ transforms. Note that $\kappa$ is related to the angle of HWP @ $\theta_2$ by $\kappa=\sin2\theta_2$. Thus, by tuning $\theta_2$, we may implement any of the filters (for $d=2$) given in \cref{Eq:OptimalFilterRhoq}. With some thought, it is easy to see that the same effect applies to every term in the convex decomposition of an input {\em mixed} density matrix.
\begin{widetext}
\begin{equation}
\begin{split}\label{eq:LocalFilter}
&\qquad\qquad\,\,\,\alpha\ket{H_{1^{\prime}}}+\beta\ket{V_{1^{\prime}}}
\xrightarrow{\text{BD}_1}\alpha\ket{H_{1^{\prime}}}\ket{h_{1^{\prime}}}+\beta\ket{V_{1^{\prime}}}\ket{v_{1^{\prime}}}\\
&\xrightarrow[\text{on  path } h]{\text{HWP}~@~\theta_2}\alpha\cos2\theta_2\ket{H_{1^{\prime}}}\ket{h_{1^{\prime}}}+\alpha\sin2\theta_2\ket{V_{1^{\prime}}}\ket{h_{1^{\prime}}}+\beta\ket{V_{1^{\prime}}}\ket{v_{1^{\prime}}}\\
&\xrightarrow[\text{on  path } v]{\text{HWP}~@~45^{\circ}}\alpha\cos2\theta_2\ket{H_{1^{\prime}}}\ket{h_{1^{\prime}}}+\alpha\sin2\theta_2\ket{V_{1^{\prime}}}\ket{h_{1^{\prime}}}+\beta\ket{H_{1^{\prime}}}\ket{v_{1^{\prime}}}\\
&\xrightarrow[{+\text{post-select path }1^{\prime\prime}}]{\text{BD}_2}\frac{\alpha\sin2\theta_2\ket{V_{1^{\prime\prime}}}+\beta\ket{H_{1^{\prime\prime}}}}{\sqrt{|\alpha|^2\sin^2 2\theta_2+|\beta|^2}}
\xrightarrow[ \text{on  path } 1^{\prime\prime}]{\text{HWP}~@~45^{\circ}}\frac{\alpha\sin2\theta_2\ket{H_{1^{\prime\prime}}}+\beta\ket{V_{1^{\prime\prime}}}}{\sqrt{|\alpha|^2\sin^2 2\theta_2+|\beta|^2}}
\end{split}
\end{equation}
\end{widetext}


\subsection{Preparation of the input state for teleportation}

\begin{figure}[h!tbp]
\includegraphics[width=0.88\columnwidth]{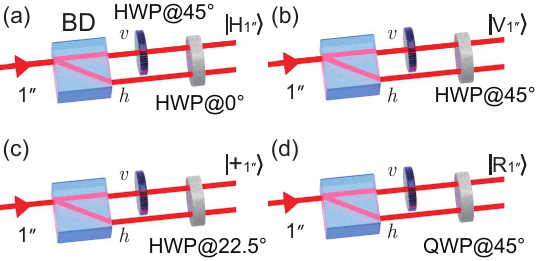}
\caption{Experimental setup (zoom-in view of the ``state preparation"  part of Fig. 2(b))  that prepares the four pure states to be teleported $\ket{\psi}=\alpha\ket{H_{1^{\prime\prime}}}+\beta\ket{V_{1^{\prime\prime}}}$. We set HWP, respectively, at 0$^{\circ}$, 45$^{\circ}$ and 22.5$^{\circ}$ to prepare $\ket{H_{1^{\prime\prime}}}$, $\ket{V_{1^{\prime\prime}}}$ and $\ket{+_{1^{\prime\prime}}}$, and QWP at 45$^{\circ}$ to prepare $\ket{R_{1^{\prime\prime}}}$. }
\label{Fig:TeleportedState}
\end{figure}

Our teleportation experiment is realized on a two-photon hybrid system. In the following, we show how this scheme works for an ideal two-photon polarization entangled state $\ket{\Phi_{2}^{+}}_{1^{\prime\prime}2}=\frac{1}{\sqrt{2}}(\ket{H_{1^{\prime\prime}}H_2}+\ket{V_{1^{\prime\prime}}V_2})$ shared between Alice and Bob. Firstly, as shown in \cref{Fig:TeleportedState}(a) and \cref{eq:TeleportedState}, the polarization-polarization entangled state $\ket{\Phi_{2}^{+}}_{1^{\prime\prime}2}$ is mapped to a two-photon path-polarization-polarization entangled Greenberger-Horne-Zeilinger state using a BD. Then, a HWP @ 45$^\circ$ placed at the spatial mode $v$ disentangles the polarization DOF of photon 1$^{\prime\prime}$ from this two-photon hybrid system. Finally, the state to be teleported is encoded in the polarization DOF of photon 1$^{\prime\prime}$ by having a HWP or a QWP set at the appropriate angle and placed across both path $v$ and $h$. The process is described as
\begin{equation}\label{eq:TeleportedState}
\begin{aligned}
	\ket{\Phi_{2}^{+}}_{1^{\prime\prime}2}&=\frac{1}{\sqrt{2}}(\ket{H_{1^{\prime\prime}}H_2}+\ket{V_{1^{\prime\prime}}V_2})\\
	&\xrightarrow{\text{BD}}\frac{1}{\sqrt{2}}(\ket{H_{1^{\prime\prime}}}\ket{H_2}\ket{h_{1^{\prime\prime}}}+\ket{V_{1^{\prime\prime}}}\ket{V_2}\ket{v_{1^{\prime\prime}}})\\
	&\xrightarrow[\text{on  path } v]{\text{HWP}~@~45^{\circ}}\frac{1}{\sqrt{2}}(\ket{H_{1^{\prime\prime}}}\ket{H_2}\ket{h_{1^{\prime\prime}}}+\ket{H_{1^{\prime\prime}}}\ket{V_2}\ket{v_{1^{\prime\prime}}})\\
	&\qquad\qquad=\ket{H_{1^{\prime\prime}}}\otimes \frac{1}{\sqrt{2}}(\ket{H_2}\ket{h_{1^{\prime\prime}}}+\ket{V_2}\ket{v_{1^{\prime\prime}}})\\
	&\xrightarrow[\text{across both paths}]{\text{HWP or QWP}}(\alpha\ket{H_{1^{\prime\prime}}}+\beta\ket{V_{1^{\prime\prime}}})\otimes\\
	&\qquad\qquad\qquad\quad\frac{1}{\sqrt{2}}(\ket{H_2}\ket{h_{1^{\prime\prime}}}+\ket{V_2}\ket{v_{1^{\prime\prime}}}).
\end{aligned}
\end{equation}
Experimentally, we choose $\ket{H_{1^{\prime\prime}}}$, $\ket{V_{1^{\prime\prime}}}$, $\ket{+_{1^{\prime\prime}}}=\frac{1}{\sqrt{2}}(\ket{H_{1^{\prime\prime}}}+\ket{V_{1^{\prime\prime}}})$ and $\ket{R_{1^{\prime\prime}}}=\frac{1}{\sqrt{2}}(\ket{H_{1^{\prime\prime}}}+{\rm i }\ket{V_{1^{\prime\prime}}})$ as the four states to be teleported. The corresponding waveplate settings are shown in ~\cref{Fig:TeleportedState}.

\subsection{Bell-state measurement (BSM) }

\begin{figure}[h!tbp]
\includegraphics[width=0.82\columnwidth]{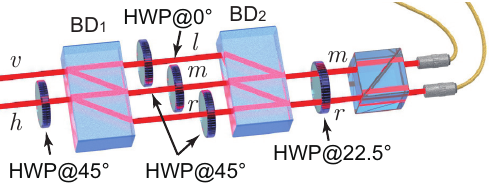}
\caption{Experimental setup (zoom-in view of the ``BSM"  part of Fig.~2b) to implement Bell-state measurement between the path and the polarization DOF of photon 1$^{\prime\prime}$. }
\label{Fig:BSM}
\end{figure}

A crucial step of the teleportation protocol is to apply a Bell-state measurement on the state to be teleported together with one half of the shared entangled resource. In our case, this amounts to applying a BSM between the polarization and path DOF of photon 1$^{\prime\prime}$. In contrast with the BSM on two photons, since this measurement is to act on two different DOFs of a single photon, all four Bell states can in principle be distinguished deterministically in a single shot. Our experimental setup for implementing this measurement is shown in \cref{Fig:BSM}, while the associated theoretical calculations are shown in ~\cref{eq:BSM}.

\begin{widetext}
\begin{equation}\label{eq:BSM}
\begin{aligned}
	&\qquad\qquad\quad\frac{1}{\sqrt{2}}(\ket{H_2}\ket{h_{1^{\prime\prime}}}+\ket{V_2}\ket{v_{1^{\prime\prime}}})\otimes(\alpha\ket{H_{1^{\prime\prime}}}+\beta\ket{V_{1^{\prime\prime}}})\\
	&\xrightarrow[\text{on  path } h]{\text{HWP}~@~45^{\circ}}\frac{1}{\sqrt{2}}(\alpha\ket{H_2}\ket{h_{1^{\prime\prime}}}\ket{V_{1^{\prime\prime}}}+\beta\ket{H_2}\ket{h_{1^{\prime\prime}}}\ket{H_{1^{\prime\prime}}}+\alpha\ket{V_2}\ket{v_{1^{\prime\prime}}}\ket{H_{1^{\prime\prime}}}+\beta\ket{V_2}\ket{v_{1^{\prime\prime}}}\ket{V_{1^{\prime\prime}}})\\
	&\xrightarrow{\text{BD}_1}\frac{1}{\sqrt{2}}(\alpha\ket{H_2}\ket{m_{1^{\prime\prime}}}\ket{V_{1^{\prime\prime}}}+\beta\ket{H_2}\ket{r_{1^{\prime\prime}}}\ket{H_{1^{\prime\prime}}}+\alpha\ket{V_2}\ket{m_{1^{\prime\prime}}}\ket{H_{1^{\prime\prime}}}+\beta\ket{V_2}\ket{l_{1^{\prime\prime}}}\ket{V_{1^{\prime\prime}}})\\
	&\xrightarrow[\text{HWP}~@~0^{\circ}\text{on  path } m]{\text{HWP}~@~45^{\circ}\text{on  path } l, r}\frac{1}{\sqrt{2}}(-\alpha\ket{H_2}\ket{m_{1^{\prime\prime}}}\ket{V_{1^{\prime\prime}}}+\beta\ket{H_2}\ket{r_{1^{\prime\prime}}}\ket{V_{1^{\prime\prime}}}+\alpha\ket{V_2}\ket{m_{1^{\prime\prime}}}\ket{H_{1^{\prime\prime}}}+\beta\ket{V_2}\ket{l_{1^{\prime\prime}}}\ket{H_{1^{\prime\prime}}})\\
	&\xrightarrow{\text{BD}_2}\frac{1}{\sqrt{2}}(-\alpha\ket{H_2}\ket{m_{1^{\prime\prime}}}\ket{V_{1^{\prime\prime}}}+\beta\ket{H_2}\ket{r_{1^{\prime\prime}}}\ket{V_{1^{\prime\prime}}}+\alpha\ket{V_2}\ket{r_{1^{\prime\prime}}}\ket{H_{1^{\prime\prime}}}+\beta\ket{V_2}\ket{m_{1^{\prime\prime}}}\ket{H_{1^{\prime\prime}}})\\
	&\xrightarrow[\text{on both paths}]{\text{HWP}~@~22.5^{\circ}}\frac{1}{2}[(\alpha\ket{H_2}+\beta\ket{V_2})\ket{m_{1^{\prime\prime}}}\ket{V_{1^{\prime\prime}}}+(\beta\ket{H_2}+\alpha\ket{V_2})\ket{r_{1^{\prime\prime}}}\ket{H_{1^{\prime\prime}}}\\
	&\qquad\qquad\quad\,\,\ +(-\alpha\ket{H_2}+\beta\ket{V_2})\ket{m_{1^{\prime\prime}}}\ket{H_{1^{\prime\prime}}}+(-\beta\ket{H_2}+\alpha\ket{V_2})\ket{r_{1^{\prime\prime}}}\ket{V_{1^{\prime\prime}}}]\\
\end{aligned}
\end{equation}
\end{widetext}

Essentially, the first four steps of the above calculation can be seen as implementing the controlled-NOT operation between the path and the polarization DOF of photon 1. The last step then amounts to implementing the Hadamard gate. As such, to complete the BSM, it suffices to measure photon 1$^{\prime\prime}$ in the complete basis $\{\ket{m_{1^{\prime\prime}}}\ket{V_{1^{\prime\prime}}}, \ket{r_{1^{\prime\prime}}}\ket{H_{1^{\prime\prime}}}, \ket{m_{1^{\prime\prime}}}\ket{H_{1^{\prime\prime}}}, \ket{R_{1^{\prime\prime}}}\ket{V_{1^{\prime\prime}}}\}$, which we achieve by putting a PBS that intersects path $m$ and $r$ after BD$_2$. In our experiment, since we are limited by the number of detectors available, we only collect the transmitted photon after PBS. This means that we only implement a partial BSM that allows us to identify $\ket{m_{1^{\prime\prime}}}\ket{H_{1^{\prime\prime}}}$ and $\ket{r_{1^{\prime\prime}}}\ket{H_{1^{\prime\prime}}}$ while being ignorant of which among the two cases $\ket{m_{1^{\prime\prime}}}\ket{V_{1^{\prime\prime}}}$ and $\ket{r_{1^{\prime\prime}}}\ket{V_{1^{\prime\prime}}}$ actually takes place. To compensate for this, we set for only about half of the experimental runs the final HWP @ 22.5$^\circ$ and the remaining runs the final HWP @ 67.5$^\circ$. Then, in these other cases, we could identify $\ket{m_{1^{\prime\prime}}}\ket{V_{1^{\prime\prime}}}$ and $\ket{r_{1^{\prime\prime}}}\ket{V_{1^{\prime\prime}}}$ while being ignorant of which among the two cases $\ket{m_{1^{\prime\prime}}}\ket{H_{1^{\prime\prime}}}$ and $\ket{r_{1^{\prime\prime}}}\ket{H_{1^{\prime\prime}}}$ actually takes place. This then allows us to cover all four possible outcomes of the BSM.

Notice that in our setup, the classical communication from Alice to Bob was only carried out after the experiment, rather than during the experiment to facilitate an active unitary correction depending on the BSM outcome. In other words, the correction unitary was realized in a post-selected manner, i.e., we applied the unitary independent of the BSM outcome and kept only those instances where our choice of unitary matched with the desired correcting unitary.

\subsection{Quantum process tomography (QPT)\\ of the teleportation channel}
\label{App:Process}

The experimental process teleporting a quantum state $\rho$ from Alice to Bob can be described by a completely-positive trace-preserving (CPTP) map $\mathcal E(\rho)$. To this end, note that we may choose $\{A_m\}_m:=\{I, X, Y, Z\}$ (where $I=\id_2$ and $X=\sigma_x,Y=\sigma_y,Z=\sigma_z$ are Pauli observables) as a basis set for linear operators acting on qubit states. The CPTP map can then be expressed as~\cite{Nielsen10}
\begin{equation} \label{eq:operatorsum}
	\mathcal E(\rho)=\sum_{m,n=1}^{4}\chi_{mn}A_n\rho A_m^\dagger,
\end{equation}
where the expansion coefficient $\chi_{mn}$ defines the $(m,n)$ element of the so-called process matrix $\chi$ (see, e.g.,~\cite{White:JOptB:2007}).

For an ideal teleportation process $\chi_{\text{id}}$, $\mathcal{E}(\rho)=\rho$, thus except $\chi_{II}=1$, all other elements of $\chi_{\text{id}}$ are 0. Experimentally, we set $q(\theta_1)$ in the range of $\frac{1}{15}$ to $\frac{10}{15}$ in steps of $\frac{1}{15}$. For each $q(\theta_1)$, we perform a teleportation experiment and reconstruct the corresponding process matrix  $\chi_{\text{exp}}$ for the shared state $\rho_{1^{\prime}2}$, $\rho_{1^{\prime\prime}2,\kappa}$ and $\rho_{1^{\prime\prime}2,\kappa^{\prime}}$, respectively.  These experimentally determined $\chi_{\text{exp}}$'s then give, via \cref{eq:operatorsum}, a full description of the corresponding teleportation channel based on the various shared entangled resource.

\begin{figure*}[h!tbp]
 \includegraphics[scale=0.8]{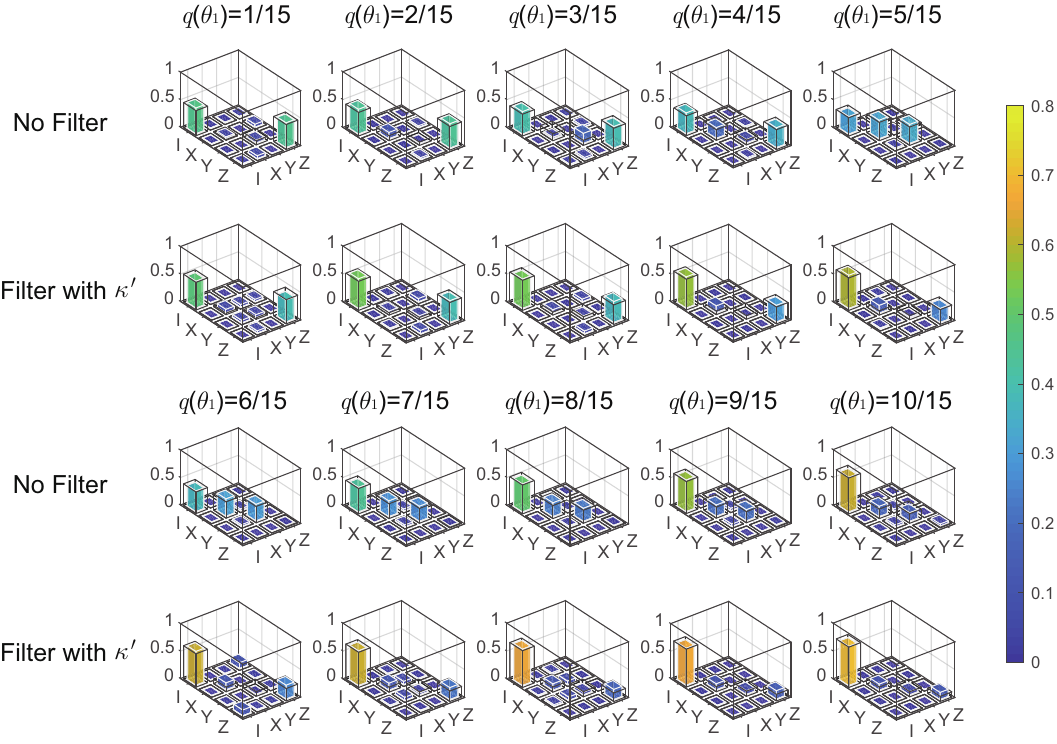}
\caption{The real parts of $\chi_{\text{exp}}$ based on the shared states $\rho_{1^{\prime}2}$ and $\rho_{1^{\prime\prime}2,\kappa^{\prime}}$. The imaginary parts are omitted here as their experimentally determined values are tiny. The wire grids represent the theoretical values of the elements.}
\label{Fig:process}
\end{figure*}

From the point of view of a process matrix, the goal of local filtering is to make the value of $\chi_{II}$ greater, which therefore results in a better teleportation fidelity. In ~\cref{Fig:process}, we show the real parts of $\chi_{\text{exp}}$ based on the shared states $\rho_{1^{\prime}2}$ and $\rho_{1^{\prime\prime}2, \kappa^{\prime}}$, which clearly illustrates that the experimentally determined $\chi_{II}$ becomes more dominant after local filtering. Notice also that $\chi_{\text{exp}}$ for $\rho_{1^{\prime\prime}2, \kappa}$ looks similar to that of $\rho_{1^{\prime\prime}2, \kappa^{\prime}}$ but with $\rho_{1^{\prime\prime}2, \kappa^{\prime}}$ giving a more pronounced increase in $\chi_{II}$. The corresponding plots of $\chi_{\text{exp}}$ for $\rho_{1^{\prime\prime}2, \kappa}$ are therefore omitted.

\subsection{Counts and other experimental results}

\begin{figure}[h!tbp]
\includegraphics[scale=0.9]{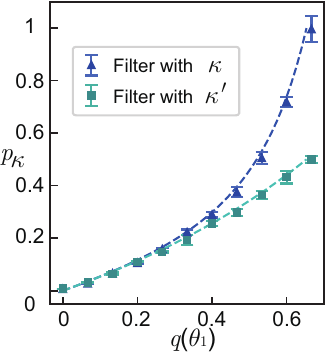}
\caption{Experimentally determined success probability of filtering $\pk$ and $\pkp$ for, respectively, filter $A_\kappa$ and $A_{\kappa'}$.}
\label{Fig:exp_data_ProbFilter}
\end{figure}

For completeness, we provide in ~\cref{tab:coincidence} the two-fold coincidence count rates of $\rho_{1^{\prime}2}$, $\rho_{1^{\prime\prime}2, \kappa}$ and $\rho_{1^{\prime\prime}2, \kappa^{\prime}}$ and in~\cref{Fig:exp_data_ProbFilter} the experimentally determined success probability of filtering.

\begin{table}[h]
\vspace{0.2cm}
\begin{tabular}{|l|l|l|l|}
\hline
$q(\theta_1)$     & $\rho_{1^{\prime}2}$     & $\rho_{1^{\prime\prime}2,\kappa}$     & $\rho_{1^{\prime\prime}2,\kappa^{\prime}}$ \\ \hline
1/15  & 7475/s  & 205/s   & 225/s   \\ \hline
2/15  & 8077/s  & 531/s   & 510/s   \\ \hline
3/15  & 8624/s  & 914/s   & 939/s   \\ \hline
4/15  & 9649/s  & 1523/s  & 1414/s  \\ \hline
5/15  & 10160/s & 2256/s  & 2026/s  \\ \hline
6/15  & 11454/s & 3316/s  & 2955/s  \\ \hline
7/15  & 13141/s & 4922/s  & 3927/s  \\ \hline
8/15  & 14683/s & 7420/s  & 5606/s  \\ \hline
9/15  & 17183/s & 12340/s & 7498/s  \\ \hline
10/15 & 20514/s & 20427/s & 10699/s \\ \hline
\end{tabular}
\caption{The two-fold coincidence count rates of $\rho_{1^{\prime}2}$, $\rho_{1^{\prime\prime}2, \kappa}$ and $\rho_{1^{\prime\prime}2, \kappa^{\prime}}$. For comparison, note that the two-fold coincidence count rate just before the photons enter the fibers are 7.3$\times$10$^4$/s.}
\label{tab:coincidence}
\end{table}

For the quantum state tomography of $\rho_{1^{\prime}2}$, $\rho_{1^{\prime\prime}2, \kappa}$ and $\rho_{1^{\prime\prime}2, \kappa^{\prime}}$, we projected the two-photon polarization state onto the $4\times4=16$ tomographically complete basis states $\{\ket{H},\ket{V},\ket{+},\ket{R} \}\otimes\{\ket{H},\ket{V},\ket{+},\ket{R} \}$. In particular, since we have only one detector on Alice's side and one on Bob's side, these 16 projections individually defines one measurement setting. For each of them, we accumulated two-fold coincidences for 1 second. Evidently, given the form of the state prepared, the counts accumulated may drastically vary from one measurement setting to another. The total number of coincidences collected for the  reconstruction of $\rho_{1^{\prime}2}$, $\rho_{1^{\prime\prime}2, \kappa}$ and $\rho_{1^{\prime\prime}2, \kappa^{\prime}}$, and hence the calculation of $F_2(\rho)$ for the various $\rho$, are shown in ~\cref{tab:Totalcounts}.

To perform quantum process tomography, we prepared separately the input states $\ket{H}$, $\ket{V}$, $\ket{+}$ and $\ket{R}$ for the teleportation channels based on the entangled states shared between Alice and Bob. After teleportation, we performed quantum state tomography on the recovered photon (photon 2 in our experiment) by projecting it onto  $\ket{H}$, $\ket{V}$, $\ket{+}$ and $\ket{R}$ respectively. In each experimental setting, we accumulated two-fold coincidences for 5 seconds except for the case of $\rho_{1^{\prime\prime}2, \kappa}$ and $\rho_{1^{\prime\prime}2, \kappa^{\prime}}$ with $q(\theta)=1/15$, in which we accumulated two-fold coincidence for 50 seconds. The total number of coincidences collected for the reconstruction of these quantum processes are shown in ~\cref{tab:Totalcounts}. In ~\cref{tab:processdata}, we show the results of state fidelity between the input state to the teleportation channel and the recovered state, as well as the corresponding results of process fidelity $\mathcal F_{p}$.

\begin{table*}[h]
\begin{tabular}{|l|l|l|l|l|l|l|}
\hline
$q(\theta_1)$     & $F_2(\rho_{1^{\prime}2})$     & $F_2(\rho_{1^{\prime\prime}2, \kappa})$     & $F_2(\rho_{ 1^{\prime\prime}2, \kappa^{\prime}})$ & $f(\rho_{1^{\prime}2})$     & $f(\rho_{1^{\prime\prime}2, \kappa})$     & $f(\rho_{1^{\prime\prime}2, \kappa^{\prime}})$ \\ \hline
 1/15& 33094~(1s)& 840~(1s)& 841~(1s)& 289024~(5s)& 71807(50s)& 72975~(50s)\\ \hline
 2/15& 35974~(1s)& 2239~(1s)& 1914~(1s)& 298344~(5s)& 19504~(5s)& 18861~(5s)\\\hline
 3/15& 38894~(1s)& 3499~(1s)& 3833~(1s)& 344394~(5s)& 31936~(5s)& 32956~(5s)\\\hline
 4/15& 42279~(1s)& 6001~(1s)& 6230~(1s)& 369119~(5s)& 55900~(5s)& 52041~(5s)\\\hline
 5/15& 45898~(1s)& 8196~(1s)& 9290~(1s)& 406941~(5s)& 83883~(5s)& 75763~(5s)\\\hline
 6/15& 51210~(1s)& 13657~(1s)& 13041~(1s)& 456755~(5s)& 169583~(5s)& 160960~(5s)\\\hline
 7/15& 56780~(1s)& 19645~(1s)& 17575~(1s)& 521988~(5s)& 172885~(5s)& 164164~(5s)\\\hline
 8/15& 65647~(1s)& 30881~(1s)& 23194~(1s)& 615689~(5s)& 295860~(5s)& 232231~(5s)\\\hline
 9/15& 77203~(1s)& 47427~(1s)& 32828~(1s)& 718538~(5s)& 443754~(5s)& 314547~(5s)\\\hline
 10/15& 91057~(1s)& 88122~(1s)& 42918~(1s)& 886182~(5s)& 836689~(5s)& 456112~(5s)\\\hline

\end{tabular}
\caption{The number of two-fold coincidences recorded for the calculation of $F_2(\rho)$ and $f(\rho)$ with $\rho_{1^{\prime}2}$, $\rho_{1^{\prime\prime}2, \kappa}$ and $\rho_{1^{\prime\prime}2, \kappa^{\prime}}$. The  data acquisition time for each measurement setting is noted in parentheses next to each entry. }
\label{tab:Totalcounts}
\end{table*}

\begin{table*}
	\centering
	\begin{tabular}{|c|c|c|c|c|c|c|c|}
		\hline
		$q(\theta_1)$  &\makecell[c]{Input\\ state}  & \makecell[c]{State fidelity\\after teleportation\\with $\rho_{1^{\prime}2}$}  & $\mathcal F_p(\rho_{1^{\prime}2})$ & \makecell[c]{State fidelity\\after teleportation\\with $\rho_{1^{\prime\prime}2, \kappa}$}  & $\mathcal F_p(\rho_{1^{\prime\prime}2, \kappa})$ &\makecell[c]{State fidelity\\after teleportation\\with $\rho_{1^{\prime\prime}2, \kappa^\prime}$}  & $\mathcal F_p(\rho_{1^{\prime\prime}2, \kappa^\prime})$ \\\hline
		\multirow{4}*{1/15} & $\ket{H}$ &  $0.921\pm0.002$ & \multirow{4}*{$0.470\pm0.003$} & $0.912\pm0.003$ & \multirow{4}*{$0.489\pm0.007$}  & $0.925\pm0.003$& \multirow{4}*{$0.486\pm0.006$} \\
		\cline{2-3}\cline{5-5}\cline{7-7}
		~ & $\ket{V}$ & $0.924\pm0.001$ &	~&  $0.871\pm0.003$& ~& $0.866\pm0.004$ &~ \\
		\cline{2-3}\cline{5-5}\cline{7-7}
		~ & $\ket{+}$ & $0.503\pm0.005$  &	~& $0.529\pm0.010$& ~& $0.541\pm0.010$ &~ \\
		\cline{2-3}\cline{5-5}\cline{7-7}
		~ & $\ket{R}$ & $0.525\pm0.005$  &	~& $0.533\pm0.010$& ~& $0.550\pm0.010$ &~\\
		\hline
		\multirow{4}*{2/15} & $\ket{H}$ &  $0.882\pm0.002$ & \multirow{4}*{$0.446\pm0.003$} & $0.893\pm0.007$& \multirow{4}*{$0.534\pm0.012$}  & $0.919\pm0.006$& \multirow{4}*{$0.539\pm0.014$} \\
		\cline{2-3}\cline{5-5}\cline{7-7}
		~ & $\ket{V}$ & $0.886\pm0.002$ &	~&  $0.903\pm0.006$& ~& $0.898\pm0.006$ &~ \\
		\cline{2-3}\cline{5-5}\cline{7-7}
		~ & $\ket{+}$ & $0.554\pm0.005$  &~&  $0.509\pm0.018$& ~& $0.554\pm0.019$ &~ \\
		\cline{2-3}\cline{5-5}\cline{7-7}
		~ & $\ket{R}$ & $0.444\pm0.004$ &	~&  $0.553\pm0.019$& ~& $0.554\pm0.020$ &~\\
		\hline
		\multirow{4}*{3/15} & $\ket{H}$ &  $0.807\pm0.002$ & \multirow{4}*{$0.404\pm0.003$} &$0.840\pm0.006$& \multirow{4}*{$0.528\pm0.010$}  & $0.879\pm0.005$& \multirow{4}*{$0.529\pm0.010$} \\
		\cline{2-3}\cline{5-5}\cline{7-7}
		~ & $\ket{V}$ & $0.777\pm0.002$  & ~ & $0.913\pm0.004$& ~& $0.932\pm0.004$ &~ \\
		\cline{2-3}\cline{5-5}\cline{7-7}
		~ & $\ket{+}$ & $0.431\pm0.004$  & ~ & $0.583\pm0.015$& ~& $0.580\pm0.015$ &~ \\
		\cline{2-3}\cline{5-5}\cline{7-7}
		~ & $\ket{R}$ & $0.578\pm0.005$  & ~ & $0.572\pm0.015$& ~& $0.574\pm0.015$ &~\\
		\hline
		\multirow{4}*{4/15} & $\ket{H}$ &  $0.752\pm0.002$ & \multirow{4}*{$0.380\pm0.003$} & $0.817\pm0.005$& \multirow{4}*{$0.546\pm0.011$}  & $0.821\pm0.005$& \multirow{4}*{$0.569\pm0.007$} \\
		\cline{2-3}\cline{5-5}\cline{7-7}
		~ & $\ket{V}$ & $0.733\pm0.002$  & ~ &  $0.858\pm0.004$& ~& $0.914\pm0.003$ &~ \\
		\cline{2-3}\cline{5-5}\cline{7-7}
		~ & $\ket{+}$ & $0.565\pm0.004$  & ~ &  $0.619\pm0.012$& ~& $0.601\pm0.013$ &~ \\
		\cline{2-3}\cline{5-5}\cline{7-7}
		~ & $\ket{R}$ & $0.451\pm0.004$  & ~ &  $0.637\pm0.013$& ~& $0.612\pm0.012$ &~\\
		\hline
		\multirow{4}*{5/15} & $\ket{H}$ &  $0.311\pm0.002$ & \multirow{4}*{$0.315\pm0.001$} & $0.780\pm0.004$& \multirow{4}*{$0.577\pm0.010$}  & $0.803\pm0.004$& \multirow{4}*{$0.597\pm0.005$} \\
		\cline{2-3}\cline{5-5}\cline{7-7}
		~ & $\ket{V}$ & $0.340\pm0.002$ &	~&  $0.838\pm0.004$& ~& $0.887\pm0.003$ &~ \\
		\cline{2-3}\cline{5-5}\cline{7-7}
		~ & $\ket{+}$ & $0.625\pm0.004$  &	~&  $0.673\pm0.010$& ~& $0.649\pm0.011$ &~ \\
		\cline{2-3}\cline{5-5}\cline{7-7}
		~ & $\ket{R}$ & $0.621\pm0.004$  &	~&  $0.683\pm0.011$& ~& $0.633\pm0.011$ &~\\
		\hline
		\multirow{4}*{6/15} & $\ket{H}$ &  $0.381\pm0.002$ & \multirow{4}*{$0.376\pm0.001$} & $0.847\pm0.002$& \multirow{4}*{$0.607\pm0.005$}  & $0.883\pm0.002$& \multirow{4}*{$0.616\pm0.004$} \\
		\cline{2-3}\cline{5-5}\cline{7-7}
		~ & $\ket{V}$ & $0.384\pm0.002$ &	~&  $0.786\pm0.003$& ~& $0.869\pm0.003$ &~ \\
		\cline{2-3}\cline{5-5}\cline{7-7}
		~ & $\ket{+}$ & $0.666\pm0.004$  &	~&  $0.712\pm0.008$& ~& $0.680\pm0.008$ &~ \\
		\cline{2-3}\cline{5-5}\cline{7-7}
		~ & $\ket{R}$ & $0.654\pm0.004$  &	~&  $0.687\pm0.008$& ~& $0.637\pm0.007$ &~\\
		\hline
		\multirow{4}*{7/15} & $\ket{H}$ &  $0.446\pm0.002$ & \multirow{4}*{$0.441\pm0.002$} & $0.661\pm0.003$& \multirow{4}*{$0.608\pm0.007$}  & $0.791\pm0.003$& \multirow{4}*{$0.624\pm0.005$} \\
		\cline{2-3}\cline{5-5}\cline{7-7}
		~ & $\ket{V}$ & $0.450\pm0.002$ &	~&  $0.789\pm0.003$& ~& $0.80\pm0.002$ &~ \\
		\cline{2-3}\cline{5-5}\cline{7-7}
		~ & $\ket{+}$ & $0.695\pm0.005$  &	~&  $0.730\pm0.008$& ~& $0.714\pm0.008$ &~ \\
		\cline{2-3}\cline{5-5}\cline{7-7}
		~ & $\ket{R}$ & $0.688\pm0.004$  &	~&  $0.724\pm0.008$& ~& $0.685\pm0.008$ &~\\
		\hline
		\multirow{4}*{8/15} & $\ket{H}$ &  $0.518\pm0.002$ & \multirow{4}*{$0.503\pm0.003$} & $0.666\pm0.003$& \multirow{4}*{$0.611\pm0.006$}  & $0.794\pm0.002$& \multirow{4}*{$0.653\pm0.004$} \\
		\cline{2-3}\cline{5-5}\cline{7-7}
		~ & $\ket{V}$ & $0.519\pm0.002$ &	~&  $0.711\pm0.002$& ~& $0.872\pm0.002$ &~ \\
		\cline{2-3}\cline{5-5}\cline{7-7}
		~ & $\ket{+}$ & $0.721\pm0.004$  &	~&  $0.757\pm0.006$& ~& $0.734\pm0.007$ &~ \\
		\cline{2-3}\cline{5-5}\cline{7-7}
		~ & $\ket{R}$ & $0.714\pm0.004$  &	~&  $0.756\pm0.006$& ~& $0.710\pm0.006$ &~\\
		\hline
		\multirow{4}*{9/15} & $\ket{H}$ &  $0.590\pm0.002$ & \multirow{4}*{$0.575\pm0.003$} & $0.671\pm0.002$& \multirow{4}*{$0.650\pm0.004$}  & $0.805\pm0.002$& \multirow{4}*{$0.672\pm0.003$} \\
		\cline{2-3}\cline{5-5}\cline{7-7}
		~ & $\ket{V}$ & $0.600\pm0.002$ &	~&  $0.703\pm0.002$& ~& $0.843\pm0.002$ &~ \\
		\cline{2-3}\cline{5-5}\cline{7-7}
		~ & $\ket{+}$ & $0.754\pm0.004$  &	~&  $0.799\pm0.005$& ~& $0.784\pm0.006$ &~ \\
		\cline{2-3}\cline{5-5}\cline{7-7}
		~ & $\ket{R}$ & $0.758\pm0.004$  &	~&  $0.794\pm0.005$& ~& $0.759\pm0.006$ &~\\
		\hline
		\multirow{4}*{10/15} & $\ket{H}$ &  $0.658\pm0.001$ & \multirow{4}*{$0.622\pm0.003$} & $0.679\pm0.001$& \multirow{4}*{$0.654\pm0.003$}  & $0.816\pm0.002$& \multirow{4}*{$0.703\pm0.003$} \\
		\cline{2-3}\cline{5-5}\cline{7-7}
		~ & $\ket{V}$ & $0.663\pm0.001$ &	~&  $0.699\pm0.001$& ~& $0.834\pm0.002$ &~ \\
		\cline{2-3}\cline{5-5}\cline{7-7}
		~ & $\ket{+}$ & $0.797\pm0.004$  &	~&  $0.789\pm0.004$& ~& $0.795\pm0.005$ &~ \\
		\cline{2-3}\cline{5-5}\cline{7-7}
		~ & $\ket{R}$ & $0.780\pm0.004$  &	~&  $0.799\pm0.004$& ~& $0.781\pm0.005$ &~\\
		\hline
	\end{tabular}
\caption{Summary of the quality of our teleportation channels based on sharing, respectively $\rho_{1^{\prime}2}$, $\rho_{1^{\prime\prime}2, \kappa}$, and $\rho_{1^{\prime\prime}2, \kappa^{\prime}}$. Note that $\rho_{1^{\prime}2}$ is the shared entangled state that was locally filtered, whereas $\rho_{1^{\prime\prime}2, \kappa}$ and $\rho_{1^{\prime\prime}2, \kappa^{\prime}}$ are the states obtained by, respectively,  applying the  local filter $A_\kappa$ and $A_{\kappa'}$. Included in the table are, for each value of $q(\theta_1)$,  the fidelity of the teleported state with respect to their input state $\{\ket{H},\ket{V},\ket{+}$, and $\ket{R}\}$, as well as the corresponding process fidelity $\mathcal F_p$.
}
\label{tab:processdata}
\end{table*}

\onecolumngrid

\subsection{Data fitting}

\begin{figure}[h!tbp]
\includegraphics[scale=.8]{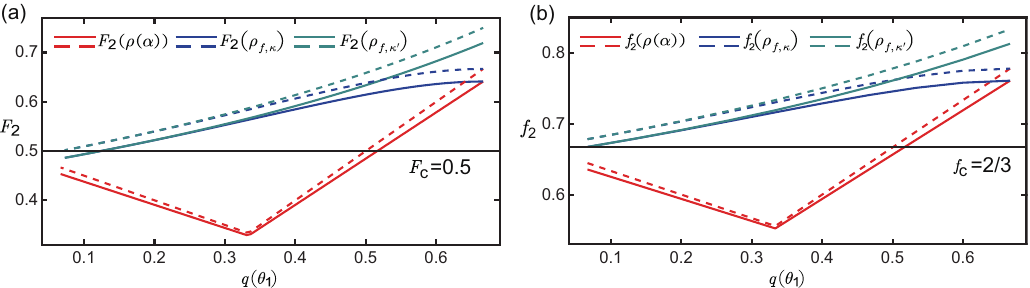}
\caption{The theoretical predictions of {\bf (a)}, the FEF $F_2$ and {\bf (b)}, the teleportation fidelity $f_2$ assuming an SPDC source described by $\rin(\alpha)$. Dashed lines represent the results of $\rin(\alpha=1)$ while solid lines represent that of $\rin(\alpha=0.954)$.
}
\label{Fig:fidelity_compare}
\end{figure}

\begin{figure}[h!tbp]
\includegraphics[scale=1.2]{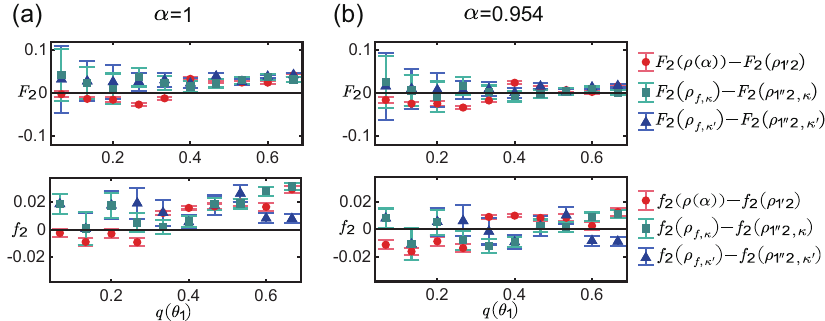}
\caption{Differences between theoretical predictions and experimental results assuming an SPDC source described by {\bf (a)}, $\rin(\alpha=1)$ and {\bf (b)} $\rin(\alpha=0.954)$.}
\label{Fig:difference}
\end{figure}

Imperfections in our experiments are mainly due to higher-order emissions in the process of spontaneous parametric down-conversion (SPDC) and slight misalignment of optical elements during the data collection. We model these imperfections by considering a noisy entangled state at stage ``1" (see Fig 2a of the main text) in the form of $\rin(\alpha)=\alpha\ket{\Psi_{2}^{+}}\bra{\Psi_{2}^{+}}+(1-\alpha) \frac{\id_{4}-\ket{\Psi_{2}^{+}}\bra{\Psi_{2}^{+}}}{4}$. In particular, $\rin(\alpha=1)$  corresponds to an ideal Bell pair $\ket{\Psi_2}$. In our experiment, we observe a $\Psi_2^+$-fidelity of $0.954\pm0.003$, which corresponds to $\rin(\alpha=0.954)$. The theoretical calculations of $F_2$ and $f_2$ with $\rin(\alpha=0.954)$ are shown as solid lines in Fig.~\ref{Fig:fidelity_compare}.  Compared with the results obtained by assuming an ideal source (dashed lines), the calculated curves for $\rin(\alpha=0.954)$ show a better fit with the experimental data. This can be seen by the difference between the theoretical predictions and the experimental results at $q(\theta)=\frac{1}{15}, \frac{2}{15}, \cdots, \frac{10}{15}$ shown in \cref{Fig:difference}. The corresponding values of $F_2$ and $f_2$ are listed in \cref{tab:predictions}.

\begin{table}[]
    \centering
\begin{tabular}{|c|l|c|c|c|c|c|c|c|c|c|c|c|}
	\hline
	\multicolumn{2}{|c|} {$q(\theta)$} & $1 / 15$ & $2 / 15$ & $3 / 15$ & $4 / 15$ & $5 / 15$ & $6 / 15$ & $7 / 15$ & $8 / 15$ & $9 / 15$ & $10 / 15$\\
	\hline \multirow{4}*{$\alpha=1$} & $F_{2}\left(\rho_{1^{\prime}2} \right)$
	 & 0.467  & 0.433 &	0.400 &	0.367 &	0.333 &	0.400 &	0.467 &	0.533 &	0.600 &	0.667
	\\
	\cline{2-12} & $F_{2}\left(\rho_{1^{\prime \prime}2, \kappa} \right)$ & 0.517 & 0.536 &	0.555 &	0.575 &	0.595 &	0.615 &	0.634 &	0.651 &	0.662 &	0.667
	\\
	\cline { 2 - 12 } & $F_{2}\left(\rho_{1^{\prime \prime}2, \kappa^{\prime}} \right)$ & 0.517 & 0.536 &	0.556 &	0.577 &	0.600 &	0.625 &	0.652 &	0.682 &	0.714 &	0.750
	\\
	\hline \multirow{4}*{$\alpha=0.954$} & $F_{2}\left(\rho_{1^{\prime}2} \right)$ & 0.453 & 0.422 &	0.391 &	0.360 &	0.328 &	0.391 &	0.453 &	0.516 &	0.579 &	0.641
	\\
	\cline { 2 - 12 } & $F_{2}\left(\rho_{1^{\prime \prime}2, \kappa} \right)$ & 0.501 & 0.518 &	0.536 &	0.555 &	0.574 &	0.593 &	0.611 &	0.626 &	0.637 &	0.641
	\\
	\cline { 2 - 12 } & $F_{2}\left(\rho_{1^{\prime \prime}2, \kappa^{\prime}} \right)$ & 0.501 & 0.518 &	0.537 &	0.557 &	0.579 &	0.602 &	0.628 &	0.655 &	0.686 &	0.719
	\\
	\hline \multirow{4}{*} {Exp.} & $F_{2}\left(\rho_{1^{\prime}2} \right)$ & 0.469(6) & 0.447(5) &	0.415(6) &	0.393(3) &	0.345(4) &	0.367(4) &	0.441(4) &	0.509(4) &	0.576(4) &	0.625(4)
	\\
	\cline { 2 - 12 } & $F_{2}\left(\rho_{1^{\prime \prime}2, \kappa} \right)$ & 0.49(8) & 0.51(5) &	0.53(4) &	0.54(2) &	0.56(2) &	0.59(1) &	0.595(7) &	0.624(7) &	0.623(3) &	0.624(4)
	\\
	\cline { 2 - 12 } & $F_{2}\left(\rho_{1^{\prime \prime}2, \kappa^{\prime}} \right)$ & 0.47(6) & 0.51(4) &	0.55(4) &	0.54(2) &	0.58(2) & 0.61(1) & 	0.63(1) &	0.652(8) &	0.677(6) &	0.719(6)
	\\

	\hline \multirow{4}{*} {$\alpha=1$} & $f_2\left(\rho_{1^{\prime} 2}\right)$ & 0.644 &	0.622 &	0.600 &	0.578 &	0.556& 	0.600 &	0.644 &	0.689 &	0.733 &	0.778 	
	\\
	\cline { 2 - 12 } & $f_2\left(\rho_{1^{\prime \prime}2, \kappa} \right)$ & 0.678 &	0.690 &	0.703 &	0.717 &	0.730 &	0.744 &	0.756 &	0.767 &	0.775 &	0.778
	\\
	\cline { 2 - 12 } & $f_2\left(\rho_{1^{\prime \prime}2, \kappa^{\prime}} \right)$ &0.678 &	0.690 &	0.704 &	0.718 &	0.733 &	0.750 &	0.768 &	0.788 &	0.810 &	0.833
	\\
	\hline \multirow{4}{*} {$\alpha=0.954$} & $f_2\left(\rho_{1^{\prime}2} \right)$ &0.635 &	0.615 &	0.594 &	0.573 &	0.552 &	0.594 &	0.636& 	0.677 &	0.719 &	0.761 	
	\\
	\cline { 2 - 12 } & $f_2\left(\rho_{1^{\prime \prime}2, \kappa} \right)$ & 0.667 	&0.679 &	0.691 	&0.703 &	0.716 	&0.729 	&0.740& 	0.751 &	0.758 &	0.761
	\\
	\cline { 2 - 12 } & $f_2\left(\rho_{1^{\prime \prime}2, \kappa^{\prime}} \right)$ & 0.667 &	0.679 &	0.691 &	0.705 &	0.719 &	0.735 &	0.752 &	0.770 &	0.791 &	0.813 	
	\\
	\hline \multirow{4}{*} {Exp.} & $f_2\left(\rho_{1^{\prime}2} \right)$ & 0.647(3) &	0.631(3) &	0.603(3) &	0.587(3) &	0.543(1) &	0.584(1) &	0.627(2) &	0.669(3) &	0.717(3) &	0.748(3) 	
	\\
	\cline { 2 - 12 } & $f_2\left(\rho_{1^{\prime \prime}2, \kappa} \right)$ & 0.659(7) 	& 0.69(1) &	0.69(1) &	0.70(1) &	0.718(9) &	0.738(5) &	0.738(7) &	0.740(6) &	0.766(4) &	0.770(3) 	
	\\
	\cline { 2 - 12 } & $f_2\left(\rho_{1^{\prime \prime}2, \kappa^{\prime}} \right)$ & 0.659(7) &	0.69(1) &	0.686(9) &	0.712(7) &	0.731(5) &	0.744(4) &	0.749(5) &	0.769(4) &	0.781(3) &	0.802(3)
	\\
	\hline
\end{tabular}
    \caption{The theoretical predictions based on $\rin(\alpha=1)$ and $\rin(\alpha=0.954)$ as well as the experimental results for the FEF $F_2$ and the teleportation fidelity $f_2$ at $q(\theta)=\frac{1}{15}, \frac{2}{15}, \cdots, \frac{10}{15}$.}
    \label{tab:predictions}
\end{table}



\end{document}